\documentclass[12pt,preprint]{aastex}         
\usepackage{graphicx}
\usepackage{epstopdf}
\def\be{\begin{equation}}
\def\ee{\end{equation}}

\def\lambar{\lambda\llap {--}}

\def\tt{\mbox{\boldmath $\theta $}}
\def\pp{\mbox{\boldmath $\phi $}}

\def\erg{\varepsilon}
\def\lambar{\lambda\llap {--}}
\def\teq#1{$\, #1\,$}                         

\def\lsim{\lower 2pt \hbox{$\, \buildrel {\scriptstyle <}\over
         {\scriptstyle \sim}\,$}}
\newcommand\gsim{\buildrel > \over \sim}
\begin{document}
\newcommand{\figureout}[3]{\psfig{figure=#1,width=5.5in,angle=#2} 
   \figcaption{#3} }

\title{Pulsar Pair Cascades in Magnetic Fields with Offset Polar Caps}

\author{Alice K. Harding\altaffilmark{1} \& Alex G. Muslimov\altaffilmark{1,2}} 
  
\altaffiltext{1}{Astrophysics Science Division,      
NASA/Goddard Space Flight Center, Greenbelt, MD 20771}

\altaffiltext{2}{Universities Space Research Association/CRESST, Columbia, MD 21044}


\begin{abstract}

Neutron star magnetic fields may have polar caps (PC) that are offset from the dipole axis, through field-line sweepback near 
the light cylinder or non-symmetric currents within the star.   The effects of such offsets on electron-positron pair cascades are 
investigated, using simple models of dipole magnetic fields with small distortions that shift the PCs by different amounts or directions.
Using a Monte Carlo pair cascade simulation, we explore the changes in the pair spectrum, multiplicity and energy flux across
the PC, as well as the trends in pair flux and pair energy flux with spin-down luminosity, $L_{\rm sd}$.  We also give an estimate of the
distribution of heating flux from returning positrons on the PC for different offsets.  We find that even modest offsets can produce significant 
increases in pair multiplicity, especially for pulsars that are near or beyond the pair death lines for centered PCs, primarily because 
of higher accelerating fields.  Pair 
spectra cover several decades in energy, with the spectral range of millisecond pulsars (MSPs) two orders of magnitude higher than for normal 
pulsars, and PC offsets allow significant extension of all spectra to lower pair energies.  We find that the total PC 
pair luminosity $L_{\rm pair}$ is proportional to $L_{\rm sd}$, with $L_{\rm pair} \sim 10^{-3} L_{\rm sd}$ for normal pulsars and 
$L_{\rm pair} \sim 10^{-2} L_{\rm sd}$ for MSPs.  Remarkably, the total PC heating luminosity for even large offsets 
increases by less than a factor of two,
even though the PC area increases by much larger factors, because most of the heating occurs near the magnetic axis.

\end{abstract} 


\pagebreak
  
\section{Introduction}

Electron-positron pair production by rotation-powered pulsars is believed to be essential to the generation of charged particles in their 
magnetospheres and winds, as well as to the generation of their coherent pulsed radio emission.  
Pairs can be produced through electromagnetic cascades initiated by accelerated particles and several sites of acceleration have been proposed.  
Near the neutron star surface above the magnetic 
polar caps (PCs), electrons may be accelerated either in vacuum gaps (Ruderman \& Sutherland 1975) or in space-charge limited flow gaps 
(Arons \& Scharlemann 1979).   The gamma-ray photons radiated by the electrons will convert to electron-positron pairs by one-photon 
pair production in the strong magnetic field near the surface, and the pairs radiate synchrotron photons that produce further generations of pairs 
(Daugherty \& Harding 1982).  In outer gap vacuum accelerators, extending to near the light cylinder, seed photons can produce pairs
through interaction with thermal X-ray photons from the hot neutron star surface to create a first generation of particles that accelerate through
the gap radiating gamma-ray photons that produce further pairs (Cheng, Ho \& Ruderman 1986).  To date, all of the simulations of pulsar pair 
cascades have assumed pure dipole magnetic fields both to derive the accelerating electric field parallel to the magnetic field and to 
calculate the spectrum and multiplicity of the pairs. 
 
The configuration of neutron star (NS) magnetic fields is currently an unsolved problem.  
The original solution for a rotating magnetic dipole in vacuum (Deutsch 1955) shows that retardation causes the field lines 
near and outside the light cylinder to sweep back in a direction opposite to that of rotation, transitioning to an electromagnetic wave at 
large distances.  The magnetic field sweepback causes the boundary of the open field volume, and the footpoints of field lines on this boundary
that define the PC, to shift relative to the magnetic axis  opposite to the rotation direction (Arendt \& Eilek 1998, Dyks \& Harding 2004). 
The amount of 
the shift depends on the magnetic field inclination to the rotation axis, with the largest offsets occurring for oblique rotators.
In the last decade, numerical solutions of force-free magnetosphere models assuming ideal MHD (${\bf E \cdot B} = 0$)  (Contopoulos et al. 1999) show 
an even larger sweepback of the magnetic field near the light cylinder as it transitions to an MHD wind (Spitkovsky 2006, Timokhin 2006), 
as currents cause significant distortions of the magnetic field.
The open field boundary and the PC of a force-free magnetosphere thus have a larger offset than the vacuum magnetosphere.
Since the vacuum magnetosphere can accelerate particles but is not loaded with charges or currents and the ideal-MHD magnetosphere has charges and currents 
but cannot accelerate particles, a real pulsar magnetosphere lies between these two solutions.  

In addition to the guaranteed distortions of the dipole field caused by retardation and currents, it is possible that there are also distortions due to 
multipolar components near the neutron star surface.   The emission from millisecond pulsars in particular shows evidence of non-dipolar field 
structure.  The  thermal X-ray pulse profiles of some millisecond pulsars (MSPs) show asymmetries which have been successfully modeled by including 
offsets of the emitting hot spot on the neutron star surface.  Since the emission likely originates from PC heating, it is argued that MSPs such as
PSR J0437-4715 (Bognanov et al. 2007) and PSR J0030+0451 (Bogdanov \& Grindlay 2009) have either offset dipoles or offset PCs.  
Modeling of the X-ray pulsations of the NSs in Low-Mass X-ray Binaries, the likely progenitors of millisecond pulsars, show possible evidence of even more extreme 
magnetic field distortions (Lamb et al, 2009), that could result from distortion of the global magnetic field by e.g. the crustal plate tectonics (Ruderman 1991). 

We have recently shown (Harding \& Muslimov 2011, HM11) that distortions of a magnetic dipole field that produce offset PCs lead to an increase in 
the multiplicity of PC pair cascades and to significant shifts of the death line for pair production of curvature radiation photons.  We found that 
even offsets that are small fractions of a neutron star radius can enable high-multiplicity pair cascades in older pulsars that with pure dipole fields would 
be pair starved.  There were a number of previous studies of how multipolar or non-dipolar neutron star fields could increase pair production 
(Ruderman \& Sutherland 1975; Arons 1983, 1997; Asseo \& Khechinashvili 2002), but all 
of these focussed on the decrease in field line radius of curvature as the primary effect. 
Arons (1997), considering the decrease in field line radius of curvature provided by an offset dipole, 
found that offsets that are large fractions of a stellar radius, (0.7 - 0.8)$R_{\rm ns}$, are required to move the pair death line enough to 
include the entire radio pulsar population.   Medin \& Lai (2010) proposed an even larger offset of 0.95 $R_{\rm ns}$, again only considering  the 
decreased field line radius of curvature.  Kantor \& Tsygan (2003, 2004) considered the space-charge limited flow (SCLF) $E_\parallel$ 
solution of a dipole field with the addition of a displaced dipole of smaller magnitude, that effectively produces narrow, bent flux tubes.
They take into account both the increase in $E_\parallel$ and decrease in radius of curvature to derive lower death lines for pair production,
and find that displacements of $0.1R_{\rm ns}$ for the smaller dipole component can significantly change the death lines for both curvature radiation (CR)
and inverse Compton scattering (ICS) pair production.  
Our study also takes into account the increase in $E_\parallel$ and particle acceleration energy, which is by far the strongest effect on 
pair multiplicity of offset dipoles or PCs.   

The study of HM11 was based on derivation of the electric field parallel to the magnetic field near the PC assuming there is free flow of 
particles from the neutron star surface, the SCLF flow boundary condition (Arons \& Scharlemann 1979).  
The resulting current density of primary charges in this case is equal to $J = \rho_{_{\rm GJ}}c$ where $\rho_{_{\rm GJ}}$ is 
the Goldreich-Julian charge density.  This current distribution differs from the current density distribution demanded by the global ideal-MHD models,
so it is presently not clear whether the steady SCLF acceleration model will be compatible with a global pulsar magnetosphere. 
However, more realistic global magnetosphere models with resistivity and dissipation are in development (Li, Spitkovsky \& Tchekhovskoy 2011).
The current density distributions of such pulsar magnetospheres with dissipation are significantly different from those with ideal-MHD 
(Kalapotharakos et al. 2011), with the ability to possibly adjust to the $J \sim \rho_{_{\rm GJ}}c$ of the PC cascades they rely on 
for their charge supply.  Furthermore, Timokhin \& Arons (2011) have performed particle-in-cell simulations of PC pair cascades and 
show that SCLF acceleration is stable for currents within a 10\% range of $\rho_{_{\rm GJ}}c$.  

In this paper we extend the study of HM11 to include a more general expression for the distorted magnetic field, treating two cases in which the caps 
from opposite magnetic poles are offset in the same (symmetric) or opposite (asymmetric) directions.  In Section 2, we give the expressions for the 
magnetic fields in these two cases, as well as the corresponding  Goldreich-Julian charge density, field line equations and parallel electric fields.  
Section 3 describes the pair cascade simulation used to derive the results for pair multiplicity and flux distributions over the PC, and 
pair spectra given in Section 4.  We also examine the changes in the distribution of PC heating luminosity with offset PCs as well as the 
total pair and heating luminosity.

\section{Magnetic Field Configuration}

Derivation of a general non-dipolar magnetic field of a neutron star using a multipole expansion gives quite complex expressions (e.g. Asseo \& Khechinashvili 2002).
Because of the mathematical complexity, this kind of model representation of the magnetic field is impractical.
Since our main purpose in this paper is to derive an expression for the accelerating electric field near the neutron star surface, we require 
field expressions from which we can derive analytic formulae for charge density for input to Poisson's Equation for the electric potential. 
We therefore propose the following heuristic model of a non-dipolar magnetic field that provides distortions leading to offset PCs. 
We introduce an azimuthal asymmetry to the field lines of an originally symmetric dipole such that the field lines over half of 
the PC have relatively smaller radius of curvature and over the other half of the PC have larger radius of curvature. 
This can be done by adding an azimuthal dependence, controlled by the parameter $a = \varepsilon \sin \phi $, 
to the magnetic colatitude of $B_{\rm r}$ and $B_{\theta}$,  as in equations (1) and (2).  
Then, from the solenoidality condition, $\nabla \cdot B = 0$,  we can derive the azimuthal component of the distorted magnetic field, which is proportional to the 
parameter $\varepsilon$.  The resulting exact expression for the azimuthal component is rather cumbersome and can be simplified by expanding the appropriate terms in 
$\varepsilon$. In our study, for the sake of simplicity, we retained only the lowest-order (linear) terms in $\varepsilon$. Note that equations (1) and (2) can be used to achieve a significant 
distortion of the field lines while keeping the dipole axis inside the offset polar cap boundary, which significantly facilitates the subsequent analytical treatment.  
Thus, equations (1) and (2) represent global open field lines of a centered dipole that are pushed and bent toward the dipole axis on one side and pushed and bent away from the dipole axis on the other side of the polar cap, respectively (see Figure 1).            
Consequently,  one side of the PC is larger and the PC is effectively shifted from the center of symmetry.   

\noindent 
{\it Case \rm A: (the magnetic field is symmetric w.r.t. the magnetic equator)}

\noindent 
In this case, in magnetic spherical polar coordinates ($\eta $, $\theta $, $\phi $) the magnetic field reads 
\be
{\bf B} \approx {B_0\over {\eta ^3}}~\left[ \hat{\bf r}~\cos \theta +{1\over 2} ~\hat{\tt}~ (1+a)~\sin \theta - \hat{\pp }~\varepsilon ~\sin \theta~\cos \theta~\sin (\phi - \phi_0) \right].
\label{B1}
\ee
\noindent 
{\it Case \rm B: (the magnetic field is asymmetric w.r.t. the magnetic equator)}

\noindent
In this case the magnetic field can be presented as 
\be
{\bf B} \approx {B_0\over {\eta ^3}}~\left[ \hat{\bf r}~\cos [\theta (1+a)] +{1\over 2} ~\hat{\tt}~ \sin [\theta (1+a)] - {1\over 2} ~\hat{\pp }~\varepsilon ~(\theta + \sin \theta~\cos \theta )~\sin (\phi - \phi_0) \right],
\label{B2}
\ee
where $B_0$ is the surface magnetic field strength at the magnetic pole,  $\eta = r/R_{\rm ns}$ is the dimensionless radial coordinate in units of stellar radius, $R_{\rm ns}$,
$a=\varepsilon ~\cos (\phi - \phi _0)$ is the parameter characterizing the distortion of polar field lines, and $\phi_0$ is the magnetic azimuthal angle defining the meridional plane of the offset PC.   We must point out that $\bf B$ as presented by formula (\ref{B2}) is solenoidal only approximately, since in the $B_{\phi }$-component we have neglected corrections of order and higher than $\sim (\varepsilon \theta )^3$ and in $B_{\rm r}$- and $B_{\theta }$-components we have neglected the corrections of order and higher than $\sim (\varepsilon \theta )^2$ and $\sim (\varepsilon \theta )^3$, respectively.    Figure 1 shows
the field lines projected in the x-z plane for cases A and B.

\noindent
The corresponding Goldreich-Julian charge density is

\noindent
{\it Case \rm A:}
\begin{eqnarray} \label{rhoGJ_A}
\rho _{\rm GJ} & \approx & - {{\Omega ~B_0}\over {2\pi c \eta ^3}}\left\{  
\left( 1-{\kappa \over {\eta ^3}} \right)\cos \chi + {3\over 2}\left[
\left( 1-{\kappa \over {2\eta ^3}}\right) \cos \phi - \right. \right. \nonumber\\ 
&  & \left. \left. \varepsilon {\kappa \over {\eta ^3}}
\sin \phi \sin (\phi - \phi _0) \right] \sin \chi \sin \theta \cos \theta \right\},
\label{rhoGJ1}
\end{eqnarray}
\noindent
{\it Case \rm B:}
\begin{eqnarray}  \label{rhoGJ_B}
\rho _{\rm GJ} & \approx & - {{\Omega ~B_0}\over {2\pi c \eta ^3}}\left\{ 
\left( 1-{\kappa \over {\eta ^3}} \right)\cos \chi + {3\over 2}\left[ 
\left( 1-{\kappa \over {2\eta ^3}} \right) \cos \theta \cos \phi + \right. \right. 
\nonumber \\
& & \left. \left. {\varepsilon \over 3} \left( 1+{{2\kappa } \over {\eta ^3}} \right) \sin \phi \sin (\phi -\phi_0)\right]\sin \chi \sin \theta \right\},
\label{rhoGJ2}
\end{eqnarray}
where $\Omega$ is the rotation rate of the NS, $\chi $ is the pulsar obliquity, $\kappa \approx 0.15~I_{45}/R_6^3$ ($I_{45} = I/10^{45}$~g$\cdot $cm$^2$, $R_6 = R_{\rm ns}/10^6$ cm, $I$ is NS moment of inertia) is the parameter accounting for the general-relativistic frame dragging (Muslimov \& Tsygan 1992).

\subsection{Electric Field in the Polar Cap Region}

\noindent 
If the magnetic axis lies in the x-z plane, $a = \varepsilon \cos \phi$ to account for an effective offset of the dipole axis in the x-z ($\phi _0 = 0$) plane and $a = \varepsilon \sin \phi $ for an effective offset of the dipole axis in the y-z plane ($\phi _0 = \pi /2$), where $0 \leq \varepsilon  < 1$.  By using formulae (\ref{B1}) and (\ref{B2}) we can get the equations of the magnetic field line, 

\noindent 
{\it Case \rm A:}
\be \label{thetaA}
\theta = \sin ^{-1}\left( \xi ~x^{{1\over 2}(1+a)} \right), 
\label{FL1}
\ee
\noindent
{\it Case \rm B:}
\be \label{thetaB}
\theta = {1 \over {1+a}}~\sin ^{-1}\left( f~\xi ~x^{{1\over 2}(1+a)} \right), 
\label{FL2}
\ee
respectively, where $x = r/R_{\rm LC}$ is the radial distance in units of the light cylinder radius, $R_{\rm LC} = c/\Omega$; $0\leq \xi \leq 1$ is the colatitude of a footpoint of polar field line normalized by the colatitude of the PC boundary. Also, in formula (\ref{FL2}) $f(a)$ is a parameter ($\lsim 1$) that can be more reliably determined via numerical tracing of the last open field line. For a rough estimate one can derive the analytic expression,
\be    
f = \left[ \sin \left( {{\pi }\over {2(1+a)}} \right) \right] ^{{1\over 2}(1+a)} , 
\label{f}
\ee
that is applicable for $a > 0$ and can be used to plot the field lines with $x,~y > 0$. The field lines with $x,~y < 0$ can be plotted by using the coordinate reflection, 
$x(y)\rightarrow -x(-y)$ and $z \rightarrow -z$. 

\noindent
The field line radii of curvature (in units of $R_{\rm LC}$) are

\noindent 
{\it Case \rm A:}
\be  
x_c = {4\over {3(1+a)\xi }}~x^{{1\over 2}(1-a)}~
{{\left[1-(3+a)(1-a)\Delta /4 \right]^ {3/2}}\over {1+a/3 - [1-2a(1+a/2)/3]\Delta /2}},
\label{xc1}
\ee
\noindent 
{\it Case \rm B:}
\be
x_c = {4\over {3f\xi }}~x^{{1\over 2}(1-a)}~
{{\left(1-3f^2\Delta /4 \right)^ {3/2}}\over {1+a/3 - f^2\Delta /2}},
\label{xc2}
\ee
respectively, where
\be 
\Delta = \xi ^2~x^{1+a}.
\ee

Neglecting the static general-relativistic corrections, we can write the accelerating electric field, assuming the boundary conditions of SCLF (cf. Harding \& Muslimov 1998; hereafter HM98).  For $\eta - 1 \gg \theta _0$, where $\theta_0 = \sin^{-1}[(\Omega R_{\rm ns}/c)^{1/2}]$ is the standard 
PC angle,

\noindent
{\it Case \rm A:}
\be   \label{Ell_1}
E_{||} \approx - {1\over 2}~{\cal E}_0~x^a~ 
\left[ 3~{{\kappa } \over {\eta ^4}}~e_{1A}~(1-\xi ^2)~\cos \chi + {3\over 8} {{\theta _0}\over {\sqrt{\eta }}}~e_{2A}~\xi (1-\xi )\sin \chi \right]~,
\label{Epar1}
\ee
\noindent
{\it Case \rm B:}
\be
E_{||} \approx - {1\over 2}~\left( {f\over {1+a}} \right) ^2~{\cal E}_0~x^a~ 
\left[ 3~{{\kappa } \over {\eta ^4}}~e_{1B}~(1-\xi^2)\cos \chi + {3\over 8} {{\theta _0}\over {\sqrt{\eta }}}~e_{2B}~\xi (1-\xi)\sin \chi \right] ,
\label{Epar2}
\ee
where ${\cal E}_0 = (\Omega R_{\rm ns}/c)^2(B_r/B)B_0$, and
\be   
e_{1\rm A} = e_{1\rm B} = 1 +{1\over 3}~a(\eta ^3 -1),
\label{e1}
\ee
\be   
e_{2\rm A} = \alpha~\left\{ (1+\xi )\cos \phi + \beta~{{\varepsilon \kappa}\over {\eta ^{(1+a)/2}}} 
\left[ 2a +{{5-3a}\over {\eta ^{(5-a)/2}}}\right] \right\} 
\label{e2A}
\ee
and
\be   
e_{2\rm B} = \alpha~{f\over {1+a}}~\left[ (1+\xi ) \cos \phi + {1\over 3}~\beta ~\varepsilon \right], 
\label{e2B}
\ee
where
\be   
\alpha = x^{a/2}\left[ 1 + 3a\left( 1- {2\over {3\eta ^{(1+a)/2}}}\right) \right]~~~{\rm and}~~~\beta  = \left[ (1+\xi)\cos \phi _0 - {4\over 5}~\xi \cos (2\phi - \phi_0) \right]. 
\label{alpha-beta}
\ee

\noindent 
Now we shall present the expressions for $E_{||}$ that are applicable at the small altitudes, $z = \eta -1 \lsim \theta _0$ (cf. formula [19] of Harding \& Muslimov 2001; hereafter HM01),

\begin{eqnarray}  \label{Ell_2}
E_{||} & \approx & - 3~{\cal E}_0~\left( \lambda \theta _0^{a(1)}\right)^2~\left[ 4\kappa~\cos \chi 
\sum_{i=1}^{\infty} G_i(k_i, z) 
{{J_0(k_i\xi)}\over {k_i^3 J_1(k_i)}}+ \right.\nonumber\\
& & \left. \mu~\theta _0~\left(\lambda \theta _0^{a(1)}\right)~\sin \chi \cos \phi ~
\sum_{i=1}^{\infty} {\tilde G}_i({\tilde k}_i, z)
{{J_1({\tilde k}_i\xi)}\over {{\tilde k}_i^3}J_2({\tilde k}_i)} \right],  
\label{Epar_small_z}
\end{eqnarray}
where $G_i = 1- exp(-\gamma _iz)$ and ${\tilde G}_i = 1- exp(-{\tilde \gamma } _iz)$; $\lambda = 1$ and $\lambda = f(1)/[1+a(1)]$ in case A and B, respectively; $\mu = 1+a(1)+[5-a(1)]\kappa /2$; $J_0$ and $J_1$ are the Bessel functions, and $k_i$ and ${\tilde k}_i$ are the positive zeros of $J_0$ and $J_1$, respectively; 
$\gamma = k_i/\lambda \theta _0^{1+a(1)}$ and ${\tilde \gamma } = {\tilde k}_i/\lambda \theta _0^{1+a(1)}$; and $a(1) = \varepsilon \cos [\phi (1) - \phi _0]$, 
where $\phi(1) = \phi(\eta = 1)$.  The expression for the parallel electric field in Eqn (\ref{Ell_2}) above corresponds to the expression for a pure dipole magnetic field given in Eqn (19) of HM01.  The main differences here are inclusion of the azimuthal asymmetry incorporated in the $a$ parameter, and the neglect of the static GR corrections. So if one takes $a = 0$, Eqn (\ref{Ell_2}) will be equivalent to Eqn (19) in HM01 with $H(1) = \delta(1) = f(1) = 1$ and $\varepsilon = 0$.

\noindent
Note that in formula (\ref{Epar_small_z}) $\phi \approx \phi (1)$ and are related via the field-line equation,
\be
\cos (\phi - \phi _0) = {{\cos [\phi (1)-\phi _0] + \varepsilon z}\over {1+\varepsilon z \cos [\phi (1)-\phi _0]}} \approx \cos [\phi(1)-\phi_0] + \varepsilon z \sin ^2 [\phi (1)-\phi _0].
\label{phi}
\ee 

We see that the magnetic field configuration of Eqns (1) and (2) produces an $E_{\parallel}$ that is significantly larger on the offset side of the PC, by the factor $\theta_0^{-2a(1)}$, than the $E_{\parallel}$ for the case of a pure dipole field.  This increase in  $E_{\parallel}$ results from the strong dependence on the field-line curvature of the Goldreich-Julian charge density in the solution of Poisson's equation for the electric potential (see also Kantor \& Tsygan 2003).  This dependence is embodied in the factor $\theta_0^2$ in the pure dipole case, and in the factor $\theta_0^{2a(1)}$ in the offset PC case, so that the PC boundary is dependent on azimuthal angle.  For the case of an offset PC, the effective PC angle, $\theta_0^{2a(1)}$, is larger on one side of the PC, producing a larger $E_{\parallel}$, and smaller on the opposite side, producing a smaller $E_{\parallel}$.

\section{Pair Cascade Simulation}

We model the electromagnetic pair cascades above a pulsar PC using a hybrid steady-state acceleration/Monte Carlo pair cascade simulation.
The first stage of the calculation, based on the self-limited acceleration model of HM98, follows a primary electron that starts 
at the neutron star surface at magnetic colatitude $\xi$ and azimuth $\phi$ with Lorentz factor $\gamma = 1$.   An initial guess is made for the 
height $h_c$ of the pair formation front (PFF) above the surface and this distance is divided into 500 equal linear steps of length $\Delta s$. 
The PFF is the height where the first pair is produced and we assume that the $E_{\parallel}$ is completely screened above this point.  This is a
good assumption for several reasons.  First,  the electric field arises due to a small imbalance
between the actual charge density and the local, rotation-induced,  Goldreich-Julian charge density. It therefore does not require much 
additional charge to short-out this field. Second, the onset of pair cascading occurs very quickly (Daugherty \& Harding 1996), 
so that the number of pairs
produced per primary particle increases rapidly over small distances.   Thus, as found in Arons (1983), 
the width of the PFF (the screening distance of the
electric field) is very small compared to other dimensions of the problem. 
The electron advances in steps of $\Delta s$, gaining energy through electrostatic acceleration and losing energy through radiative losses. 
The PFF results from pairs produced by $\gamma$-rays of energy 
$\epsilon_{\rm min}$ (in units of $mc^2$) radiated by particles of
energy $\gamma_{\rm min}$.  The height of the PFF is then:
\be \label{Sc}
S_c = {\rm min}[S_a(\gamma_{\rm min}) + S_p(\epsilon_{\rm min})]
\ee
where $S_a(\gamma_{\rm min})$ is the distance required to accelerate
the particle until it can radiate a photon of energy $\epsilon_{\rm
min}$, and $S_p(\epsilon_{\rm min})$ is the pair attenuation length of 
the photon. The acceleration distance, $S_a(\gamma_{\rm min})$, is 
determined by first integrating the equation of motion of the particle 
to determine its energy as a function of its pathlength $s$: 
\be \label{dgamma}
c{d\gamma\over ds} = {e\over mc}E_{\parallel} - \dot\gamma_{_{\rm IC}} - 
\dot\gamma_{_{\rm CR}},
\ee
where $E_{\parallel}$ is the electric field induced parallel to the
magnetic field, $\dot\gamma_{_{\rm IC}}$ and $\dot\gamma_{_{\rm CR}}$
are the loss rates for ICS and CR.  
Expressions for $\dot\gamma_{_{\rm IC}}$ and $\dot\gamma_{_{\rm CR}}$ and details 
of their derivation are given in HM98.  
The pair production attenuation length of photons radiated by the particle through either ICS or CR 
is then determined.  The attenuation length $S_p(\epsilon)$, defined to be the path length 
over which the optical depth is unity, is given by
\be \label{tau}
   \tau(\epsilon)\; =\;\int_0^{S_p(\epsilon)} 
T_{\rm pp}(\theta_{\rm kB},\, \epsilon )\, ds\; =\; 1\quad ,       
\ee
where \teq{ds} is the pathlength differential along the photon momentum
vector \teq{{\bf k}}, $T_{\rm pp}$ is the attenuation coefficient for 
one-photon pair production and $\theta_{\rm kB}$ is the angle between 
\teq{{\bf k}} and the local magnetic field direction.

The method we use to compute the electron-positron pair attenuation
length of the photons, $S_p(\epsilon)$, has been described in detail
in Harding, Baring \& Gonthier (1997). Using equation (\ref{tau}), $S_p(\epsilon)$ is 
computed by integrating the pair production attenuation coefficient of 
the photon along its path through the dipole field. The photon is
assumed to pair produce at the point where $\tau(\epsilon) = 1$, 
and $S_p(\epsilon)$ is then set to that path length.  The two main
inputs needed are the energy of the photon and its angle to
the magnetic field, $\theta_{\rm kB}$. The energies of the radiated
photons for the purpose of computing the PFF are taken to be
(following HM98) $\epsilon_p = 13\epsilon_{\rm CR}$ for $B < 0.1\,B_{\rm cr}$, and 
$\epsilon_p = 4\epsilon_{\rm CR}/3$ for $B > 0.1\,B_{\rm cr}$, where 
$\epsilon_{\rm CR} = (3/2)(\lambar/\rho_c)\gamma^3$ is the critical CR 
energy, $\lambar = \lambda/2\pi$ is the electron classical radius, $B_{\rm cr} = 4.413 \times 10^{13}$ G, and 
$\rho_c = x_c R_{\rm LC}$ is the field line radius of curvature 
(Eqn [\ref{xc1}] and [\ref{xc2}]).   For 
the angle of the radiated photons at the emission point, 
we assume that $\theta_{\rm kB}
= 2/\gamma$ for Compton scattered photons (Dermer 1990) and 
$\theta_{\rm kB} = 0$ for CR photons.  The CR and ICS photons are assumed to be 
completely polarized in the parallel mode.  Due to the 
curvature of the field lines, $\theta_{\rm kB}$ will grow as the
photon propagates, roughly as $\sin\theta_{\rm kB}
\sim s/\rho_c$.  To evaluate the pair production attenuation coefficient
at each point along the path of the photon, we Lorentz-transform the 
photon energy and local magnetic field to the frame in which the 
photon propagates perpendicular to the local field.  This is the 
center-of momentum frame for the created pair, where the attenuation
has its simplest form and the photon energy is $\epsilon_{_{\rm CM}} = 
\epsilon\sin\theta_{\rm kB}$.  The one-photon pair attenuation coefficient is
considered in two regimes.  For $B < 0.1\,B_{\rm cr}$, photon pair
produce far above threshold, where the asymptotic expression in the
limit of large numbers of kinematically available pair
Landau states (Tsai \& Erber 1974, Daugherty \& Harding 1983 [DH83]) can be used:
\be   \label{eq:ppasymp}
   T^{\rm pp}_{\parallel,\perp} = {1\over 2}{\alpha\over \lambar} B'
   \Lambda_{\parallel,\perp}(\chi),
 \ee

\be 
   \Lambda_{\parallel, \perp}(\chi) \approx \left\{
     \begin{array}{lr} 
     (0.31, 0.15)\, \exp \mbox{\Large $(-{4\over 3\chi})$} & \chi \ll 1 
     \\ \\
     (0.72, 0.48) \, \chi^{-1/3} & \chi \gg 1
     \end{array} \right.          \label{eq:ppratlim}
\ee
where $\chi \equiv f\,\epsilon_{_{\rm CM}}/2B'$, $\alpha$ is the 
fine-structure constant, $B' = B/B_{\rm cr}$ is the dipole
field strength at point $s$ along the photon path.  
When $B > 0.1\,B_{\rm cr}$, pair production will occur near threshold,
where the above expression is not accurate.  We thus include the
factor $f = 1+0.42\erg_{_{\rm CM}}^{-2.7}$ in $\chi$, introduced
by DH83, as an approximation to the near-threshold attenuation 
coefficient.  
In this paper we compute the attenuation length averaged over photon 
polarization. 

The path of each input photon is traced through the magnetic field, 
accumulating the survival probability for pair production, $P_{\rm surv}$:
\begin{equation}
   P_{\rm surv}(s) = \exp\Bigl\{-\tau(s)\Bigr\} 
\end{equation}

Each photon may pair produce or escape,
based on a combination of the running survival probability for 
pair production.  For this simulation, we have neglected photon splitting
since pair production dominates the photon attenuation in magnetic fields $B \lsim 
10^{13}$ G (Baring \& Harding 2001).  

The ``first guess" value 
of $h_c$, and thus also of $S_c$, sets the initial acceleration
length. equation (\ref{dgamma}) is integrated in discrete steps
upward from the starting point, computing $E_{\parallel}$ from
equation (\ref{Ell_2}), 
$\dot\gamma_{_{\rm IC}}$ and $\dot\gamma_{_{\rm CR}}$ at each step.  
At each step, the pair attenuation lengths, $S_p(\epsilon)$, of CR test photons radiated by the particle of energy
$\gamma(s)$ are computed from equation (\ref{tau}).  The pair
attenuation length, and thus the value of $S_c$, also computed 
at every step, is initially infinite, because the energy of the photons 
is small, but decreases with distance as the energy of the radiated
photons increases.  Although the photon attenuation length continues to
decrease, the particle acceleration length is increasing and $S_c$ has 
a minimum.  This minimum value of $h_c = (S_c)_{\rm min}$ is 
adopted as the new height of the
electron PFF.  The electron is
accelerated again with the new value of $h_c$, producing a new PFF at the
next value of $h_c = (S_c)_{\rm min}$.  
The process is repeated, converging to a self-consistent value of $h_c$.  

Once the PFF height is established, the full pair cascade spectrum is computed 
using the Monte Carlo simulation, adopting the electron Lorentz factor as a function 
of height, $\gamma (s)$, from the self-consistent PFF calculation for the acceleration 
region from $s = 0$ to $s = h_c$.   Above the PFF, it is assumed that $E_{\parallel} = 0$
due to pair screening (HM01), so that the electrons lose energy every step due to 
CR and ICS.  The step size for $s > h_c$ is limited to less than the distance over which 
they would lose 10\% of their energy.  Each step, the electron radiates a CR spectrum,
divided into a number of logarithmically spaced bins.  The number of CR photons in each 
energy bin, $n_{\rm CR}$, is determined by the energy loss rate and average energy in that bin.
A representative photon from each bin having the average bin energy is followed
to its pair conversion or escape point, as described above.  The pair produced by the 
photon, or the escaping photon number, is then weighted by  $n_{\rm CR}$.  
If a photon pair produces, the total energy, Landau state and parallel
momentum of the electron and positron are determined.  Each member of the pair is 
assumed to have half the energy and the same direction of the parent photon.
Each member of the pair occupying an excited state emits a sequence of cyclotron or
synchrotron photons.  The method used to simulate the cyclotron/synchrotron emission
is similar to that of Daugherty \& Harding (1996).  If the particle Landau level
is larger than 20, the high-energy limit of the quantum synchrotron transition rate
(Sokolov \& Ternov 1968) is used, in which case we assume that the photons are
emitted perpendicular to the magnetic field in the particle rest frame (high-energy limit).
When the particle Landau level is smaller than 20, the exact QED cyclotron transition 
rate (Harding \& Preece 1987) is used, in which case the angles of the emitted photons
are sampled from a distribution.  In both cases, the emitted photon polarizations are
sampled from the corresponding polarization distributions.  Each emitted photon is
propagated through the magnetic field from its emission point until it 
pair produces or escapes.  The cyclotron/synchrotron emission sequence continues 
until each particle reaches the ground state.  By use of a recursive routine that is called upon the emission 
of each photon, we can follow an arbitrary number of pair generations.  
The cascade continues until all photons from each branch have escaped. 
The cascade pairs are binned in energy and the magnetic colatitude $\xi$ and azimuth $\phi$ on the  
PC of the initial test electron.  

\section{Results}

\subsection{Pair Multiplicity and Flux Distributions}

For a given set of input parameters that include pulsar period $P$, surface magnetic field $B_0$,  
inclination angle $\chi$, and offset degree $\varepsilon$, direction $\phi_0$ and symmetry (case A or B),
the cascade calculation produces the pair spectrum per primary electron as a function of $\xi$ and surface $\phi$,
$dN_+(E, \xi, \phi)/dE$, where we will express the pair energy $E$ in units of $mc^2$.  The pair multiplicity, the 
number of pairs per primary electron, as a function of $\xi$ and $\phi$ across the PC is
\be \label{M+}
M_+(\xi, \phi) = \int_{E_{\rm min}}^{E_{\rm max}} {dN_+(E, \xi, \phi)\over dE} dE
\ee
while the total pair multiplicity from each PC is
\be \label{M+PC}
M^{\rm PC}_+ = \int_0^{2\pi} d\phi \int_0^1 M_+(\xi, \phi)\, \theta_{\rm PC}^2\,R_{\rm ns}^2\,\xi d\xi.
\ee
Here, the PC angle varies with $\phi$,
\be \label{sinthetaPC}
\sin\theta_{\rm PC} = (R_{\rm ns}\Omega/c)^{(1+a)/2}.
\ee
which comes from setting $\xi = 1$ and $r = R_{\rm ns}$ in Eqn (\ref{thetaA}).
The total flux of pairs emerging from each PC is 
\be \label{N+}
\dot N_+(\xi, \phi) = \int_{E_{\rm min}}^{E_{\rm max}} {d\dot N_+(E, \xi, \phi)\over dE} dE
\ee
where 
\be \label{dotdNdE}
{d\dot N_+(E, \xi, \phi)\over dE} = {dN_+(E, \xi, \phi)\over dE}\, \dot N_p(\xi, \phi)
\ee
and where $\dot N_p(\xi, \phi)$ is the primary flux over the PC,
\be \label{Np}
\dot N_p(\xi, \phi) = n_{\rm GJ}\,c \pi R_{\rm ns}^2 \sin^2\theta_{\rm PC}.
\ee
where $n_{\rm GJ} = \rho_{\rm GJ}(\eta=1)/ e$ is the Goldreich-Julian number density at the NS surface.
In Eqn (\ref{Np}), we use the Goldreich-Julian charge density given in Eqns (\ref{rhoGJ_A}) and (\ref{rhoGJ_B}).
Similarly, the energy flux of pairs from each PC is
\be
\dot E_+(\xi, \phi) = \int_{E_{\rm min}}^{E_{\rm max}} {d\dot N_+(E, \xi, \phi)\over dE} E\,dE
\ee

We have computed $dN_+(E, \xi, \phi)/dE$ for a range of parameters to explore the distributions of $M_+(\xi, \phi)$, 
$\dot N_+(\xi, \phi)$ and $\dot E_+(\xi, \phi)$ across the PC.  
Our results are dependent on neutron star equation of state (E0S) through the $E_\parallel$ dependence on the frame-dragging, 
specifically through the $\kappa$ parameter, and on the neutron star radius, through the PC angle. There have been a number of 
measurements of neutron star mass for MSPs that give values well above the canonical $1.4 M_{\sun}$ and as high as $2 M_{\sun}$ 
(Demorest et al. 2010).  We have therefore assumed different EoS for normal (non-recycled) pulsars and MSPs.  
For non-recycled pulsars, we take $M_{\rm ns} = 1.45\,M_{\sun}$ and radius $R_{\rm ns} = 10$ km, which give a moment of inertia $I = 1.13 
\times 10^{45}\,\rm g\,cm^2$  (Lattimer \& Prakash 2007).   We take a rotating NS model (Friedman et al. 1986) for MSPs with  
$M_{\rm ns} = 2.15\,M_{\sun}$,  $R_{\rm ns} = 9.9$ km and $I = 1.56 \times 10^{45}\,\rm g\,cm^2$.

Examples of the pair multiplicity and pair flux distributions over the PC are shown in Figures 2 and 3, for a pulsar 
with $P = 0.3$ s and $B_0 = 3 \times 10^{12}$ G, and for Case B with $\phi_0 = \pi /2$ for different values of the 
offset degree $\varepsilon$.  A pulsar with these parameter values lies above but near the CR pair death line for $\varepsilon 
= 0$, in which case the cascade produces a modest pair multiplicity.  As shown in Figure 2, the maximum pair multiplicity $M_+(\xi, \phi)$ 
occurs for $\xi \sim 0.5$ and is nearly symmetric in $\phi$, with only a small asymmetry caused by the $\phi$ 
dependence of $E_\parallel$.  As $\varepsilon$ increases, the peak pair multiplicity increases but only in a small region 
of the PC toward the offset, in this case at $\phi = 270^0$, while the multiplicity decreases in the other side of the PC in 
the $\phi = 90^0$ direction.  For larger offsets, the region of high multiplicity expands as the PC boundary expands in the 
direction of the offset.   The distribution of pair flux for this same case, shown in Figure 3, is somewhat different from the 
distribution of pair multiplicity.   For $\varepsilon = 0$, $\dot E_+(\xi, \phi)$ peaks in a small ring around the center of the PC and decreases outside the ring,
again with a small asymmetry due to the $\phi$ dependence of $E_\parallel$.  The concentration of $\dot E_+(\xi, \phi)$ closer to 
the PC center is due to the fact that the Lorentz factor of the primary electron is maximum near the magnetic axis
because the decrease in radius of curvature of the field lines raises the altitude of the PFF (see Figures 4 and 5 of HM98).   
For $\varepsilon > 0$, the pair energy flux increases and becomes asymmetric, but with a greater concentration toward the PC 
center than the multiplicity.

Figure 4 shows the distribution of pair multiplicity $M_+(\xi, \phi)$ for the case of a millisecond pulsar with $P = 2$ ms, $B_0 = 
5 \times 10^8$ G, and for Case A with $\phi_0 = \pi/2$ for $\varepsilon$ between 0 and 0.6.  These parameters place the pulsar 
below the CR pair death line with very small pair multiplicity for $\varepsilon = 0$, although the distribution of $M_+(\xi, \phi)$ can be
seen to show a more pronounced asymmetry toward the `favorably curved' field lines at $\cos\phi > 0$ than in the case of the 
longer period pulsar shown in Figure 2.  
For pulsars with shorter periods, the second term in $E_\parallel$ depending on $\theta_0 \sin\chi\cos\phi$ is relatively larger 
since the value of $\theta_0$ is larger.  
As $\varepsilon$ increases, the peak pair multiplicity again increases in a small region of the PC toward the offset, but with a stronger
asymmetry toward the $\cos\phi > 0$ side of the PC.  For millisecond periods there is also less increase of ellipticity of the PC with 
increasing $\varepsilon$.  The period dependence of the PC ellipticity can easily be seen from theh expression Eqn (\ref{sinthetaPC}) for the PC angle.
If we define the elongation factor $e= \theta_{\rm PC}(\phi=3\pi/2)/\theta_0 = \theta_0^{-\varepsilon}$, then $e$ is higher for smaller $\theta_0$.

Figure 5 illustrates the effect of inclination angle on the distribution of pair multiplicity and pair flux for a millisecond pulsar with 
$P = 3$ ms, $B_0 = 4 \times 10^8$ G, and for Case A with $\phi_0 = \pi/2$ and $\varepsilon = 0.4$.  $M_+(\xi, \phi)$ peaks in 
roughly the same region of the PC for $\chi = 30^\circ$ and $\chi = 60^\circ$, but the asymmetry toward favorably curved field lines is 
is stronger in the case of $\chi = 60^\circ$.  Both the maximum multiplicity and flux are higher for $\chi = 60^\circ$.

The values of peak multiplicity,  $M_+(\xi = 0.5, \phi = 270^\circ)$, for the radio pulsar population in the $P$-$\dot P$ plane 
are shown in the contour plots in Figures 6 - 8.  Figure 6 shows contours of $\log(M_+)$ in the case of a centered PC ($\varepsilon = 0$),
where we have assumed the different neutron star EoS for normal (non-recycled) and MSPs described above.  It can be seen that 
$M_+$ is high for short periods and higher $\dot P$, reaching a maximum of a few times $10^4$, and drops sharply towards the 
death line even on a log scale.  About half of the normal pulsar population and most of the MSP population lies below the death 
line, a result that has been noted in a number of previous papers (e.g. Arons 1997, Zhang, Harding \& Muslimov  2000, Hibschman \& Arons 2001, 
Harding \& Muslimov 2002).  In Figure 7 and 8, we show contours of $\log(M_+)$ for offset PCs with $\varepsilon = 0.4$ having symmetric
(case A) and asymmetric (case B) offsets.  In both symmetric and asymmetric offsets, the regions of high pair multiplicity spread
to lower $\dot P$ and longer periods.  The increases in $M_+$ are most dramatic for that part of the population near and below the 
pair death line having very low $M_+$ in the centered PC case.  The pair death line thus moves down through nearly  the entire population of both 
recycled and non-recycled pulsars for an offset of $\varepsilon = 0.4$.  For pulsars with high $\dot P$, $M_+$ changes much less with even a large
offset, with multiplicity saturating below $\sim 5 \times 10^4$.  This saturation of $M_+$, noted by HM11, is caused by several effects.
Even in the centered PC case, when the magnetic field increases above $\sim 5 \times 10^{12}$ G pairs are increasingly produced in the 
low-lying Landau states (Baring \& Harding 2001), which results in fewer synchrotron photons and fewer generations in the cascade.
For increasing PC offsets, the resulting increase in accelerating field and particle energy initially produces higher $M_+$ pair cascades, but
as the increasingly energetic particles produce higher energy CR photons, the pairs are produced at smaller angles to the magnetic field.
This results in higher average pair energies with pairs in lower Landau states, further reduces the number of synchrotron photons.
There are some significant differences in the $\log(M_+)$ contours
for cases A and B.  The death line in the normal pulsar population is lower for case B and overall values of $M_+$ are a bit higher.  For MSPs, the 
case B death line moves up for very short periods.

The death lines for both $\varepsilon = 0$ and $\varepsilon = 0.4$ are flatter for the shorter period MSPs and the death lines at all periods are 
flatter as for $\varepsilon = 0.4$.  As noted by HM11, particle acceleration in pulsars that require a large fraction of the open field voltage to
produce pairs is limited by CR reaction.  In this regime, the electron Lorentz factors to not reach the full voltage drop between the neutron star 
and the PFF, but instead reach a lower steady-state Lorentz factor as the acceleration is balanced by the CR loss rate.  Even for centered
PCs, acceleration in MSPs operates completely in the CR reaction limit (Luo et al. 2000, Harding \& al. 2002).

\subsection{Pair Spectra}

The total spectrum of cascade pairs, integrated over the whole PC, is
\be \label{N+PC}
{d\dot N_+(E)\over dE} = \int_0^{2\pi} d\phi \int_0^1 {d\dot N_+(E, \xi, \phi)\over dE}\, \theta_{\rm PC}^2\,R_{\rm ns}^2\,\xi d\xi.
\ee
Figure 9 shows the integrated PC pair spectra for $B_0 = 3 \times {12}$ G and a range of periods  typical of non-recycled pulsars,
for centered and offset PCs.  The spectra exhibit turnovers at low and high energies that depend on period.  For centered PCs, the low-energy turnovers
occur around Lorentz factor $E = 100$ and the spectra extend up to $E \sim 10^5$, with the range increasing for shorter $P$.  For offset PCs, the 
spectra extend to lower energies, by about a decade for $\varepsilon = 0.4$.  Thus the total PC pair flux increases because pairs can be produced at lower 
energies.  Pair spectra for parameters typical of MSPs are shown in Figure 10.  For both centered and offset PCs, the MSP pair energies are much higher
that those of normal pulsars, by a factor of about 100.  The spectra extend from a low-energy turnover at $E \sim 10^4$ to a high energy cutoff around 
$10^7$ for $\varepsilon = 0$, and from $E \sim 2-3 \times 10^3$ to $\sim 10^7$ for $\varepsilon = 0.6$.  Thus the highest-energy pairs are nearly as energetic
as the primary electrons!  The large difference between normal and MSP pair spectra is due to the difference in field strengths.  In lower fields, photons 
must pair produce at higher energies, decreasing the relative $M_+$, even though the larger PC sizes of MSPs give smaller curvature radii.  But as is the 
case for normal pulsars, the increase in $M_+$ for offset PCs results from lower possible pair energies.

\subsection{Total Polar Cap Pair Flux and Luminosity}

The pair flux emerging from each PC is
\be \label{Ndotpair}
\dot N_{\rm pair} = \int_{E_{\rm min}}^{E_{\rm max}} {d\dot N_+(E)\over dE} dE
\ee
which we examine as a function of pulsar spin-down luminosity, $L_{\rm sd}$.  Figure 11 plots the PC pair flux vs. $L_{\rm sd}$ for the non-recycled 
pulsar population for different degrees of offset.  The lines are least-squares fits to the points for each case, resulting in the following 
approximate expressions:
\begin{eqnarray}
\dot N_{\rm pair} = &1.6 \times 10^{34}\,{\rm s^{-1}}\, L_{{\rm sd},35}^{1.07}, & \varepsilon = 0 \\
& 5.5 \times 10^{34}\,{\rm s^{-1}}\,L_{{\rm sd},35}^{0.98}, & \varepsilon = 0.2 \\
& 3.5 \times 10^{35}\,{\rm s^{-1}}\, L_{{\rm sd},35}^{0.87}, & \varepsilon = 0.4
\end{eqnarray}
where $L_{{\rm sd},35} \equiv L_{\rm sd}/10^{35}\,\rm erg\,s^{-1}$.  The pair flux is closely proportional to spin-down luminosity for 
$\varepsilon = 0$, with the dependence somewhat flattening for offset PCs due to a saturation at high $L_{\rm sd}$.  Figure 12 shows the 
equivalent plot for MSPs with the least-squares fits giving the following relations between $\dot N_{\rm pair}$ and $L_{\rm sd}$:
\begin{eqnarray}
\dot N_{\rm pair} = &9.5 \times 10^{33}\,{\rm s^{-1}}\, L_{{\rm sd},35}^{0.85}, & \varepsilon = 0 \\
& 8.5 \times 10^{33}\,{\rm s^{-1}}\,L_{{\rm sd},35}^{0.91}, & \varepsilon = 0.2 \\
& 3.1 \times 10^{34}\,{\rm s^{-1}}\, L_{{\rm sd},35}^{0.68}, & \varepsilon = 0.6
\end{eqnarray}

The total pair luminosity from each PC is 
\be \label{Lpair}
L_{\rm pair} = \int_{E_{\rm min}}^{E_{\rm max}} {d\dot N_+(E)\over dE} E\,dE
\ee
Plots of the PC pair luminosity vs. $L_{\rm sd}$ for non-recycled pulsars are shown in Figure 13, with the least-squared fits yielding the 
relations:
\begin{eqnarray}
L_{\rm pair} = & 2.0 \times 10^{31}\,{\rm erg\,s^{-1}}\, L_{{\rm sd},35}^{0.96}, & \varepsilon = 0 \\
& 2.7  \times 10^{31}\,{\rm erg\,s^{-1}}\,L_{{\rm sd},35}^{0.90}, & \varepsilon = 0.2 \\
& 8.2 \times 10^{31}\,{\rm erg\,s^{-1}}\, L_{{\rm sd},35}^{0.81}, & \varepsilon = 0.4.
\end{eqnarray}
The pair luminosity seems to also be roughly proportional to spin-down luminosity, with the dependence flattening for higher $\varepsilon$, 
with the efficiency for pair luminosity being $\eta_{\rm pair} = L_{\rm pair}/L_{\rm sd} \sim 2 - 8 \times 10^{-4}$.

The $L_{\rm pair}$ vs. $L_{\rm sd}$ for MSPs is shown in Figure 14, with the corresponding relations from the least-squares fits:
\begin{eqnarray}
L_{\rm pair} = & 3.1 \times 10^{32}\,{\rm erg\,s^{-1}}\, L_{{\rm sd},35}^{0.86}, & \varepsilon = 0 \\
& 3.2  \times 10^{32}\,{\rm erg\,s^{-1}}\,L_{{\rm sd},35}^{0.86}, & \varepsilon = 0.2 \\
& 7.8 \times 10^{32}\,{\rm erg\,s^{-1}}\, L_{{\rm sd},35}^{0.69}, & \varepsilon = 0.6.
\end{eqnarray}
The pair luminosity efficiency for MSPs is much higher than for non-recycled pulsars, with $\eta_{\rm pair} \sim 3 - 8 \times 10^{-3}$,
about an order of magnitude higher.   Since MSPs have much lower surface magnetic field strengths, they need to use a higher fraction 
of the total open-field line voltage to produce pairs, resulting in a higher $\eta_{\rm pair}$.  MSPs also have higher $\gamma$-ray luminosity
efficiencies (Abdo et al. 2009), because the primary particles which radiate the highest energy emission reach higher energies before pairs
screen the accelerating field.   

\subsection{Polar Cap Heating}

Pair cascades from particle accelerators near the PC will inevitably produce backflowing particles in the process of pair screening of the 
electric field.  The backflowing particles will accelerate through the same potential drop as the primary particles and deposit this energy
on the neutron star surface, increasing the surface temperature of the PC.  The radiation from the hot PC is predicted to be in the X-ray band
(Arons 1981, HM01), and observations have provided evidence for emission from hot PCs, especially in the case of middle-aged and 
millisecond pulsars (Zavlin 2007).  The predicted X-ray luminosity from heated PCs in the case of SCLF accelerators with centered PCs (HM01) 
roughly agrees with observed luminosities.  With the possibility of offset PCs increasing the pair multiplicity near the neutron star surface, it is 
important to check that the returning positron luminosity does not overheat the PC and violate the X-ray luminosity constraints.  For this estimate, we
do not perform a detailed screening calculation with pair dynamics to derive the screening scale length, as in HM01.  Rather we will compute 
the maximum returning positron fraction which will give a conservative limit on the positron heating luminosity.

\noindent
The maximum fraction of returning positrons from the PFF can be estimated as (see formula [33] in HM01) 
\be
{{\rho _{+}(z_{_{\rm PFF}})}\over {\rho _{\rm GJ}(z_{_{\rm PFF}})}} = {1\over 2}\left[1-{{\rho (z_{_{\rm PFF}})}\over {\rho _{\rm GJ}(z_{_{\rm PFF}})}}\right],
\label{rho_plus}
\ee 
where $z_{_{\rm PFF}}=\eta _{_{\rm PFF}} -1$ is the PFF dimensionless altitude and $\rho (z_{_{\rm PFF}})$, the primary charge density at the PFF,
is derived by inserting $\eta = 1$ inside the curly brackets in Eqns (\ref{rhoGJ_A}) and (\ref{rhoGJ_B}) and $\eta = \eta _{_{\rm PFF}}$ outside the brackets.

The heating flux over the PC from returning positrons can then be estimated as
\be 
{dL_{+}(\xi, \phi )\over dS} \approx c \, \rho _{+}(z_{_{\rm PFF}}, \xi , \phi )\, \gamma (z_{_{\rm PFF}}, \xi, \phi ) 
\ee
where $\gamma (z_{_{\rm PFF}}, \xi, \phi )$ is the primary electron energy at the PFF on the field line with surface $\xi$ and $\phi$.   
The reason that we use the the primary energy instead of the potential drop between the surface and the PFF to estimate the energy of the returning 
positrons at the surface is that for some cases the particles become radiation-reaction limited before reaching the PFF, so that the potential drop
would be an overestimate.   

Figure 15  shows the distribution of positron heating flux over the PC, $dL_{+}(\xi, \phi )/dS$, for the case of a non-recycled pulsar with $P = 0.1$ s and
$B_0 = 3 \times 10^{12}$ G, for different $\varepsilon$ values, while Figure 16 shows results for a MSP with $P = 2$ ms and $B = 2 \times 10^9$ G.
It is evident that most of the heating flux is concentrated at the center of the PC near the magnetic axis near $\xi = 0$, 
the contrast between the PC center and edges
being about a factor of ten.  The returning positron fraction is maximum both near $\xi = 0$ and $\xi = 1$, where the PFF is at the highest altitude.  However, the $\gamma (z_{_{\rm PFF}}, \xi, \phi )$ is maximum near $\xi = 0$, causing a maximum heating flux near the PC center.  
For $\varepsilon > 0$, most of the 
offset side of the PC is heated with lower flux than near the PC center.  There is also an asymmetry such that the favorably curved field lines have higher
heating flux, also because of higher returning positron energies.  The substantially lower heating in the outer part of the offset side of the PC, 
which has a much larger area, 
keeps the increase in total PC heating due to the offset to a minimum.  In fact in Figure 15 and 16, the physical size of the hottest area is comparable 
for all values of $\varepsilon$ that we examined.  Thus the most intensely heated area is several times smaller than the canonical PC size, even for large 
$\varepsilon$ values.  In the case of the MSP in Figure 16, this effectively smaller heated spot is shifted from the PC center by a smaller 
fraction of the canonical PC radius than the shift of the whole PC.   For the non-recycled pulsar in Figure 15 the heated area remains nearly 
centered while the whole PC is shifted by a larger fraction.

The total heating power (of precipitating positrons) can be estimated as (cf. formula [61] of HM01)  
\be
L_{+} \approx  c \int _{{\cal S}(z_{_{\rm PFF}})} \, \rho _{+}(z_{_{\rm PFF}}, \xi , \phi )\, \gamma (z_{_{\rm PFF}}, \xi, \phi ) dS,
\label{}
\ee
where the integration is over the area of a sphere cut by the polar flux tube at the radial distance $\eta _{_{\rm PFF}}$.  Figure 17 shows the 
dependence of $L_+$, as a fraction of spin down luminosity, on pulsar characteristic age, $\tau = P/2\dot P$, for different period and offsets.  
Comparing to the results in Fig. 7 of HM01 for the $\varepsilon = 0$ case, the $L_+/L_{\rm sd}$ here is higher by factors of 2 - 5 since the values we have 
given here are the upper limits on positron heating rate, while the HM01 results are from numerical computations of the $E_\parallel$ screening.
Our estimates here nevertheless serve to test the heating of offset PCs against the observational constraints.  
In the case of both normal pulsars and MSPs, the increase in heating for $\varepsilon = 0.2$ is very minimal.  Even for $\varepsilon = 0.4$, the heating 
luminosity increases by no more than a factor of 2, and for MSPs there is almost no increase except in cases where there are different degrees of 
pair screening for increasing offsets (e.g. $P = 2$ ms at large ages).

Thermal components have been detected from both middle-aged pulsars and MSPs in the soft X-ray band.  In the case of middle-aged pulsars like 
PSR B1055-52 and PSR B06556+14, typically both hot and cool thermal components appear in the spectra.  The hot components, with luminosity $L_h$, 
may be due to PC 
heating and the cool components from neutron star cooling, but in any case hot components provide a limit to any theoretical heating of the PC.  
For these pulsars, the PC heating efficiency $L_h/L_{\rm sd} = 1.4 \times 10^{-3}$ for B0656+14, and $5 \times 10^{-4}$ for B1055-52 (DeLuca et al. 2005), which 
are within a factor of 2 of the our estimated maximum heating efficiencies.  In the case of MSPs, several have measured thermal components, 
including PSR J0437-4715, J2124-3358 and J0030+0451.  Since these pulsars are too old to have significant cooling, their thermal emission 
is very likely from PC heating.  Their measured PC heating efficiencies, $L_h/L_{\rm sd} \sim 6 \times 10^{-4}$ (Zavlin 2007) are a factor of 2-3 
lower than the maximum estimated efficiencies in Figure 17. 

We have examined the change in magnetic flux over the PC as $\varepsilon$ increases.  For an offset of $\varepsilon = 0.2$ for a 0.1 s pulsar, 
the magnetic flux on the offset side of the PC is typically a factor of 2 higher than for the centered PC, while on the side opposite the offset the flux is a factor of 3 lower, and for larger offsets the flux ratio can be several orders of magnitude.  For MSPs, the contrast in flux is much smaller with ratios 
of only a factor of 4-5 in flux across the PC for offset as high as $\varepsilon = 0.6$. 

\section{Discussion}

We have investigated the effects of offset polar caps on electron acceleration and pair cascades near the neutron star surface, including
the cases where the PCs of the two hemispheres are offset symmetrically by the same amount or asymmetrically.   The asymmetric offset
case would apply to the PCs shifted by retardation and/or currents of the global magnetosphere, while the symmetric case could apply
to neutron stars with some interior current distortions that produce multipolar components near the surface.  The asymmetric PC offsets 
have now been shown to be standard in pulsar magnetosphere global geometry.  The PCs in the retarded vacuum magnetosphere are 
shifted in a direction opposite to that of the rotation by an amount that varies from a maximum of 20\% of the standard PC radius at 
$\chi = 90^\circ$ to no shift for $\chi = 0^\circ$ (Dyks \& Harding 2004).  
In terms of our offset parameter $\varepsilon$, the ratio of offset to standard PC angle is
$\theta_{\rm PC}/\theta_0 \simeq \theta_0^{-\varepsilon}$, so that for the vacuum case $\varepsilon_{\rm vac} \simeq 0.03 - 0.1$, where the large 
values of $\varepsilon_{\rm vac}$ apply to MSPs with large $\theta_0$.  In the force-free
ideal-MHD magnetosphere, the offset percentages range from 45\% at $\chi = 90^\circ$ to 30\%  at $\chi = 30^\circ$ (Bai \& Spitkovsky 2010), 
giving a range $\varepsilon_{\rm IMHD} \simeq 0.09 - 0.2$, again with the larger values of $\varepsilon_{\rm IMHD}$ applying to MSPs.

From our calculations, we find that the pair multiplicity and pair flux for offset PCs is distributed very asymmetrically over the PC, 
with higher values on the side of the PC toward the offset.  The regions of peak pair flux and multiplicity occur for $\varepsilon = 0$ at about 
half the PC radius, symmetrically around the magnetic axis.  As $\varepsilon$ increases, small regions of higher peak flux and multiplicity appear  
toward the offset and grow larger with increasing offset.  The increase in peak multiplicity moves the pair death lines downward in 
$P$-$\dot P$ space, to encompass nearly the entire radio pulsar population for $\varepsilon = 0.4$.  Since such offset values are higher than 
those resulting from retardation and currents in pulsar magnetosphere models, they would likely require interior currents that produce
large-scale non-dipolar fields near the neutron star surface.  Examination of the pair spectra reveal that the increase in multiplicity in pulsars 
with offset PCs comes from the extension of the spectra to lower pair energies.  

The discovery of pulsed gamma-ray emission from a large number of millisecond pulsars by the  Large Area Telescope (LAT) on the 
$Fermi$ Gamma-Ray Space Telescope (Abdo et al. 2009) has revealed light curves that are best modeled by 
narrow radiation gaps in the outer magnetosphere (Venter et al. 2009, Abdo et al. 2010).  Such narrow gaps require screening of the 
accelerating electric field over most of the magnetosphere by a pair multiplicity that is orders of magnitude larger than standard models 
of PC acceleration with no offsets are able to produce (Harding \& Muslimov 2002).  Since PC offsets that are only small fractions of a 
stellar radius result in large increases in pair multiplicity, and there is evidence for such offsets in MSPs, offset PCs may be a viable 
explanation for the larger-than-predicted pair activity in MSPs.  

We find that the integrated pair flux and luminosity from each PC is roughly proportional to the pulsar spin down luminosity, both 
in the case of non-recycled pulsars and MSPs with MSPs being somewhat more efficient in converting spin-down to pair  luminosity.
For pulsars with offset PCs, both pair flux and luminosity increase with offset but to a greater degree for low spin down power. 
Due to a saturation of the pair multiplicity, pulsars with high $L_{\rm sd}$ have only modest increases in PC pair flux even for large 
offsets.  We estimate that the Crab pulsar produces a pair flux from each PC of about $10^{38}\,\rm pairs\,s^{-1}$ in the case of no
offset, $\sim 2 \times 10^{38}\,\rm pairs\,s^{-1}$ for $\varepsilon = 0.2$ and  $\sim 5 \times 10^{38}\,\rm pairs\,s^{-1}$ for $\varepsilon = 0.4$.
The flux from both PCs is still more than an order of magnitude smaller than the pair flux required to account for the radiation from the nebula, 
which is estimated to be about $\gsim 4 \times 10^{40}\,\rm pairs\,s^{-1}$ (DeJager et al. 1996).  The trend of pair flux and pair luminosity 
proportional to spin down power may have interesting implications for explaining observed trends in pulsar pulsed and un-pulsed X-ray 
luminosity that are also proportional to spin down power (Vink et al. 2011).  Such a trend contrasts that of observed $\gamma$-ray 
luminosities which are proportional to $L_{\rm sd}^{1/2}$ (Thompson et al. 1997, Abdo et al. 2010b).  The $\gamma$-ray luminosity trend
can be understood if the emission is produced by the primary PC current (proportional to $L_{\rm sd}^{1/2}$) accelerated by a constant 
voltage of about $10^{13}$ V (Harding 1981, Arons 1996).  The X-ray luminosity trend may be understood if both the pulsed emission 
from the magnetosphere and the unpulsed emission from the pulsar wind nebula (PWN) are produced by pairs.  It is generally believed that pairs
are producing the emission from PWNe, and a number of high-energy emission models argue that secondary electron-positron pairs 
produce the pulsed X rays high in the magnetosphere through synchrotron and/or ICS 
(Cheng et al. 1986, Romani 1996, Takata et al. 2007, Hirotani 2008, Harding et al. 2008).

The asymmetric distribution of particle acceleration and pair multiplicity that results from offset PCs should produce asymmetries in 
observed pulsar emission.  The structure and energetics of the proposed slot gaps (SG) that form between the boundary of the open 
magnetic field and the upward curving PFF (Arons \& Scharlemann 1979), and can accelerate particles to high altitude 
(Muslimov \& Harding 2004), could be strongly affected by a an offset PC.   The particle Lorentz factor
$\gamma$ in the SG, which is expected to reach curvature radiation-reaction limit such that $\gamma \propto E_{\parallel}^{1/4}$, 
will be larger on one side of the PC, producing CR emission power proportional to $E_\parallel$ that is larger than for a dipole field.  
In older pulsars that do not produce enough pair multiplicity and screening to form SGs in centered PCs, SGs may form on only one
side of an offset PC.  An azimuthal asymmetry of both the radiation power and width of the SG 
would change both the $\gamma$-ray luminosity and the sharp $\gamma$-ray peaks that are due to caustics formed by radiation from 
trailing edge field lines (Dyks \& Rudak 2003).   Since the PC offsets that result from retardation and currents in pulsar magnetospheres 
occur toward the trailing side of 
the PC, the $\gamma$-ray peaks should be enhanced relative to the off-peak emission that is due to emission along the leading-edge 
field lines.  The ratio of $E_\parallel$, and thus $\gamma$-ray flux, between trailing and leading edges of the SG is predicted to be 
$\sim \theta_0^{-4\varepsilon} (1+\varepsilon)/(1-\varepsilon)$, which can be larger than an order of magnitude for pulsars having short periods.
SG model light curves assuming the emission asymmetry predicted in vacuum or non-ideal MHD magnetospheres fit the 
pulsar light curves measured by $Fermi$ significantly better that those of symmetric SGs (Harding et al. 2011, DeCesar et al. 2011).
The asymmetry in pair flux in offset PCs should produce asymmetries in pulsar radio emission if the radio flux is 
proportional somehow to the pair flux.  In fact, evidence for such asymmetries in pulsar radio emission have been observed in the form of partial cone 
emission (Mitra \& Rankin et al. 2011).
In a future study, we will explore the effect of an offset PCs on the acceleration of particles in the SG at high altitudes and the change to 
the shapes of $\gamma$-ray light curves.

\acknowledgments  
We acknowledge support from the NASA Astrophysics Theory and Fundamental Physics Program, the $Fermi$ Guest Investigator Program and 
the Universities Space Research Association.  AKH also thanks the Aspen Center for Physics.

\newpage
\begin{figure}
\includegraphics[width=140mm]{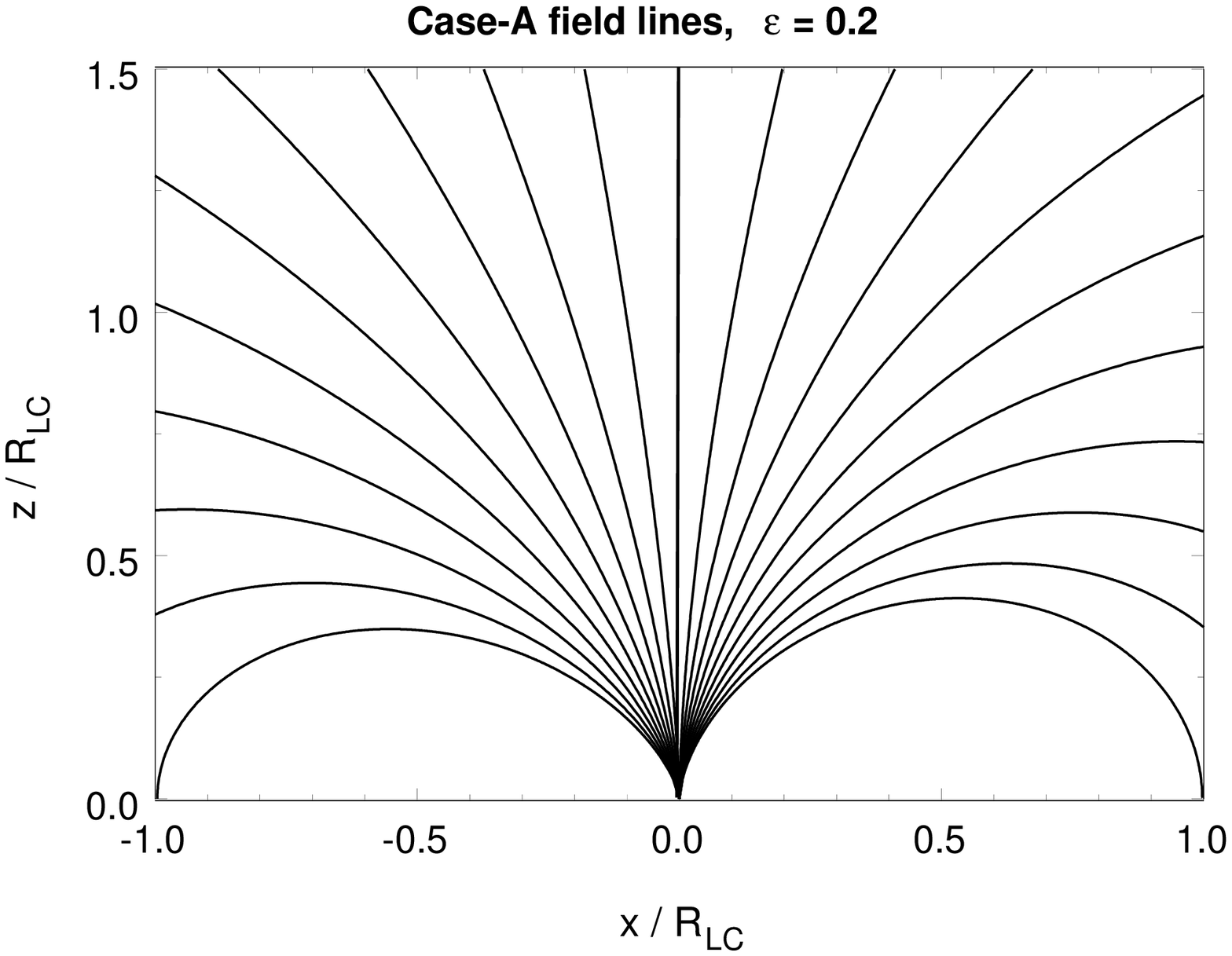}
\includegraphics[width=140mm]{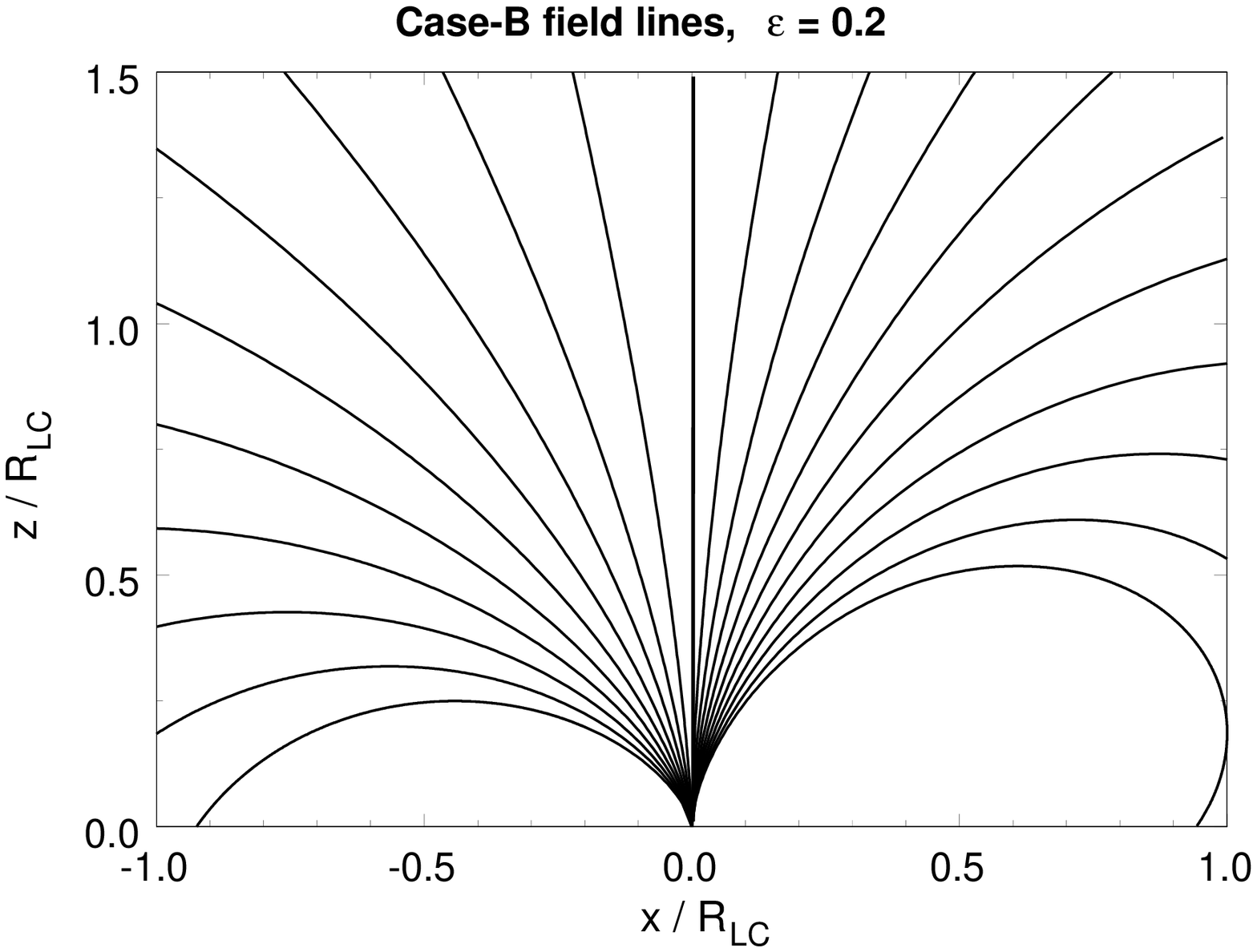}
\caption{Field lines of distorted magnetic dipole having an offset polar cap in the x-z plane and offset parameter $\varepsilon = 0.2$ for the symmetric (case A, top) 
and asymmetric (case B, bottom) polar cap offsets.}    
\end{figure}

\newpage 
\begin{figure}
\hspace{-2.0cm}
\includegraphics[width=210mm]{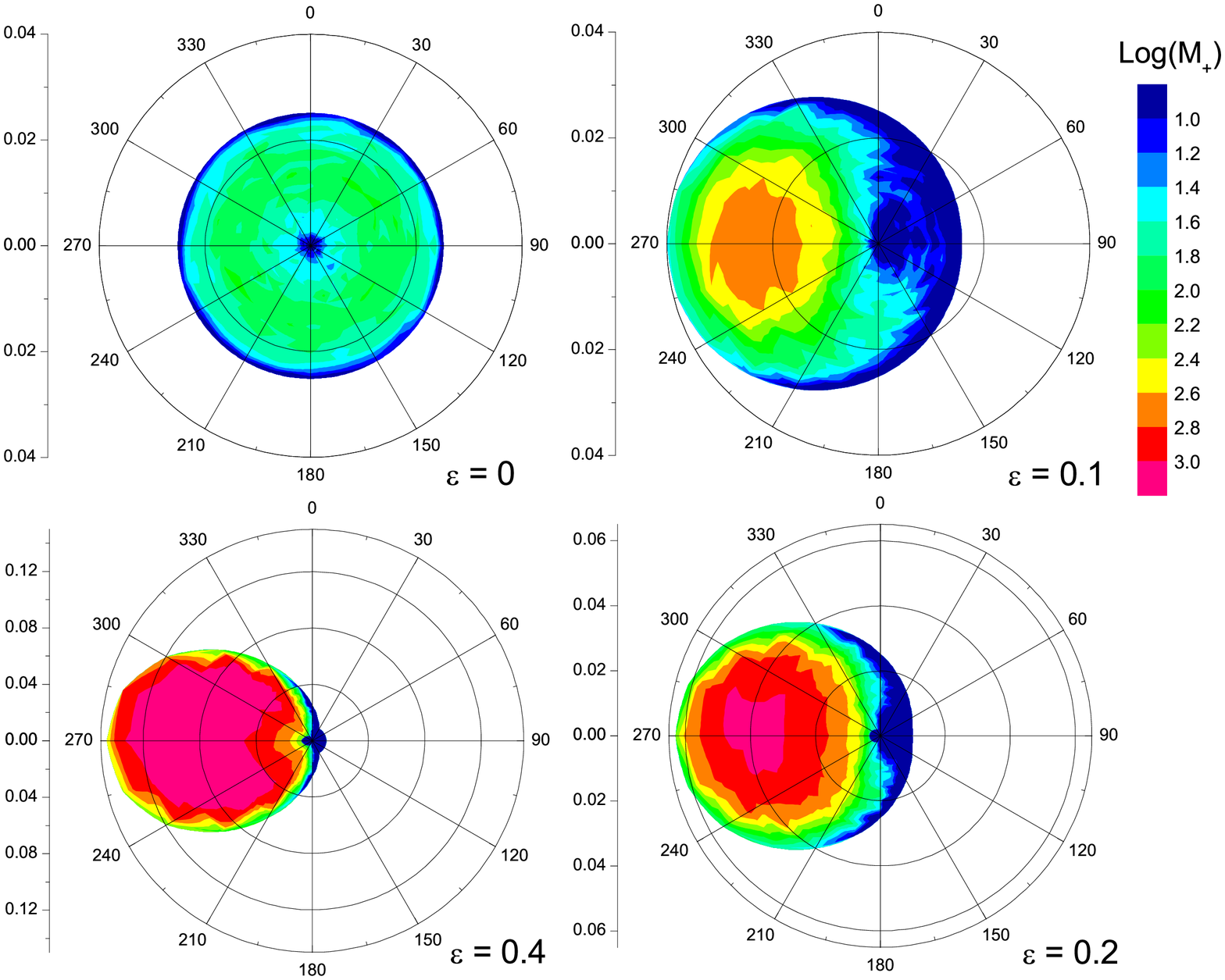}
\caption{Contours of log of pair multiplicity as a function of radial distance from the magnetic axis (in units of neutron star radius) and magnetic azimuth for 
$P = 0.3$ s, $B_0 = 3 \times 10^{12}$ G and $\chi = 60^\circ$,
for different values of offset parameter $\varepsilon$.  }    
\end{figure}

\newpage 
\begin{figure}
\hspace{-2.5cm}
\includegraphics[width=210mm]{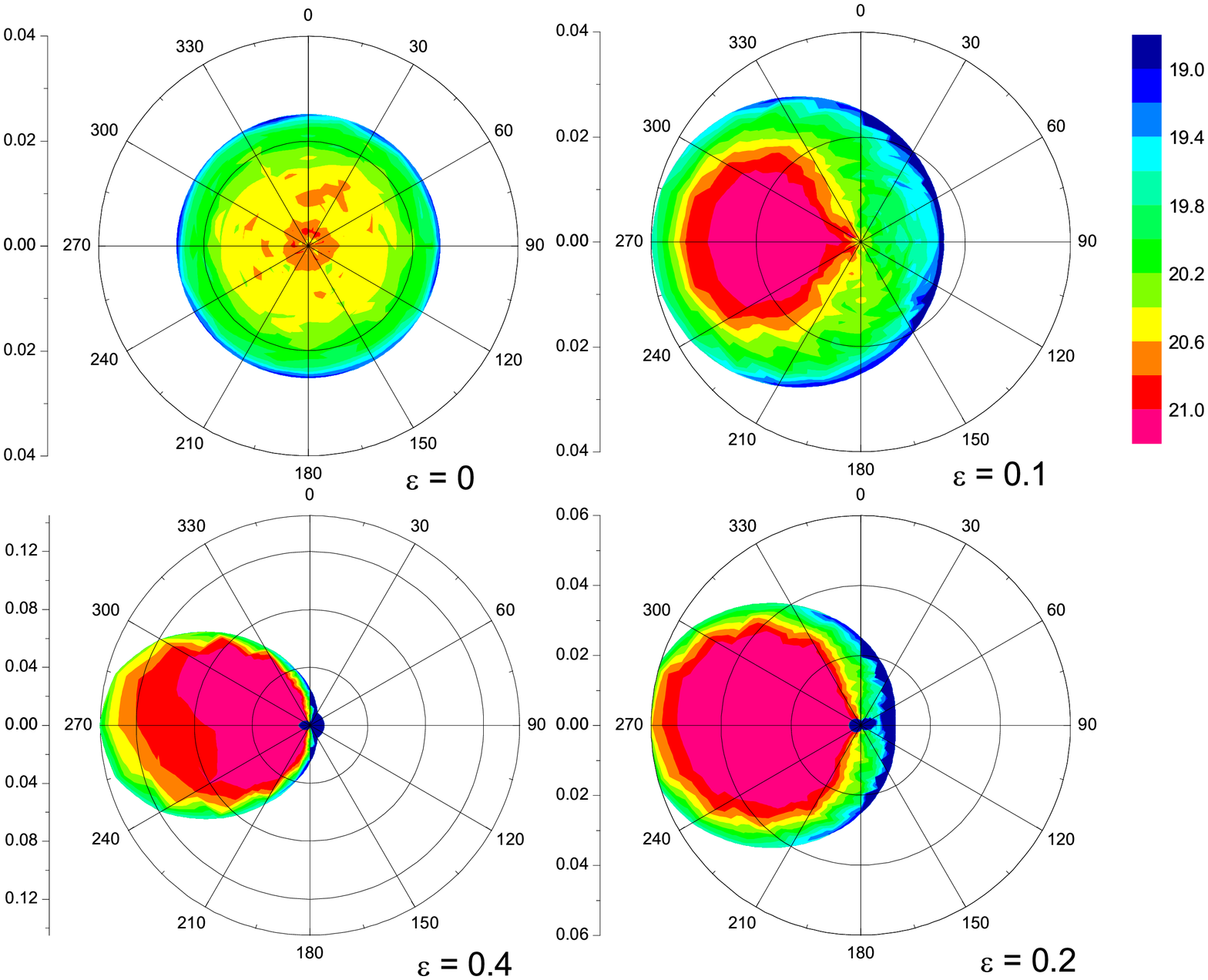}
\caption{Contours of log of pair energy flux (in units of $\rm erg\,s^{-1}\,cm^{-2}$) as a function of radial distance from the magnetic axis 
(in units of neutron star radius) and magnetic azimuth for 
$P = 0.3$ s, $B_0 = 3 \times 10^{12}$ G and $\chi = 60^\circ$, for different values of offset parameter $\varepsilon$. }    
\end{figure}

\newpage 
\begin{figure}
\hspace{-2.0cm}
\includegraphics[width=210mm]{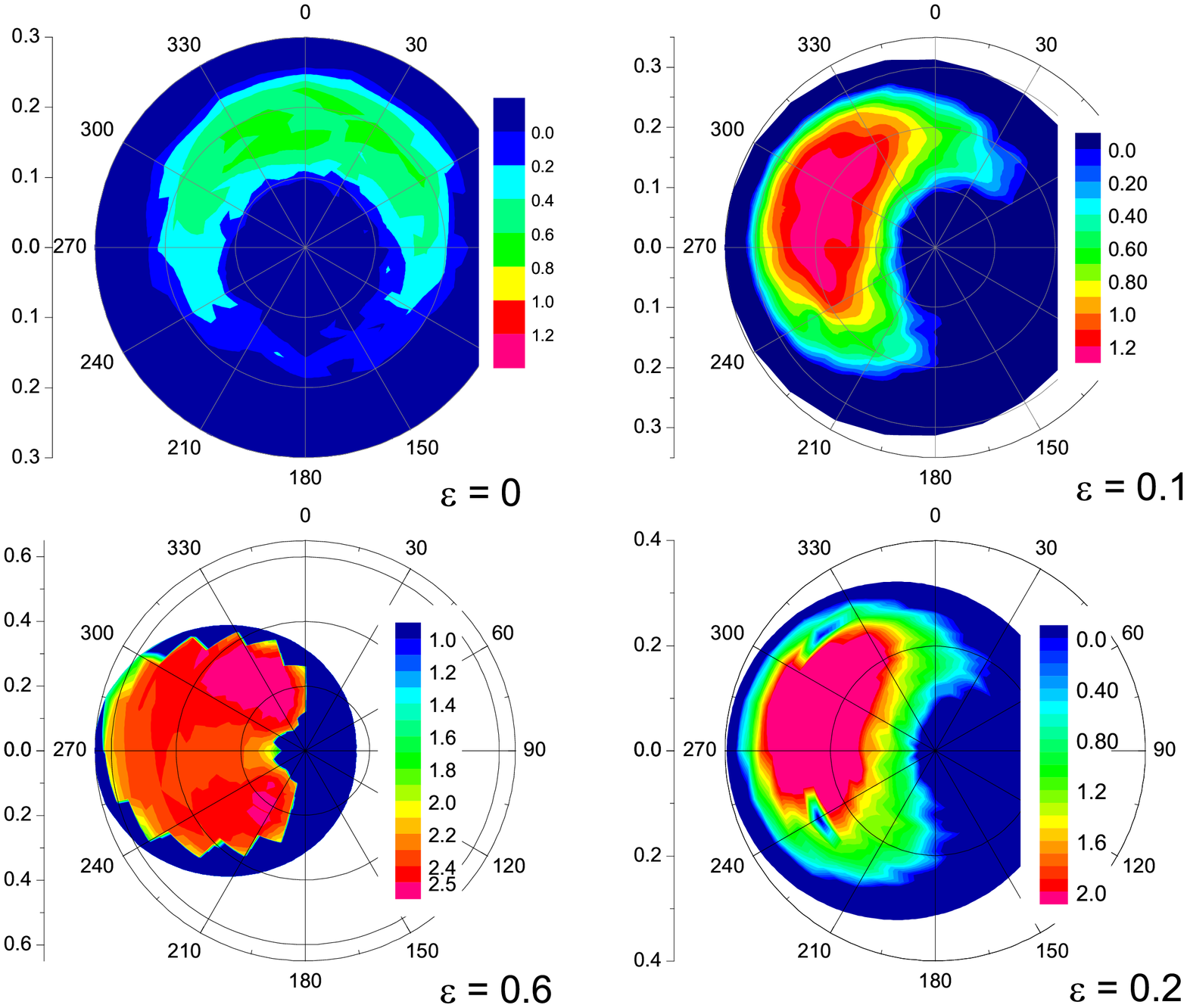}
\caption{Contours of log of pair multiplicity as a function of radial distance from the magnetic axis (in units of neutron star radius) and magnetic azimuth 
for $P = 2$ ms, $B_0 = 5 \times 10^8$ G and $\chi = 45^\circ$,
for different values of offset parameter $\varepsilon$. }    
\end{figure}

\newpage 
\begin{figure}
\hspace{-2.0cm}
\includegraphics[width=210mm]{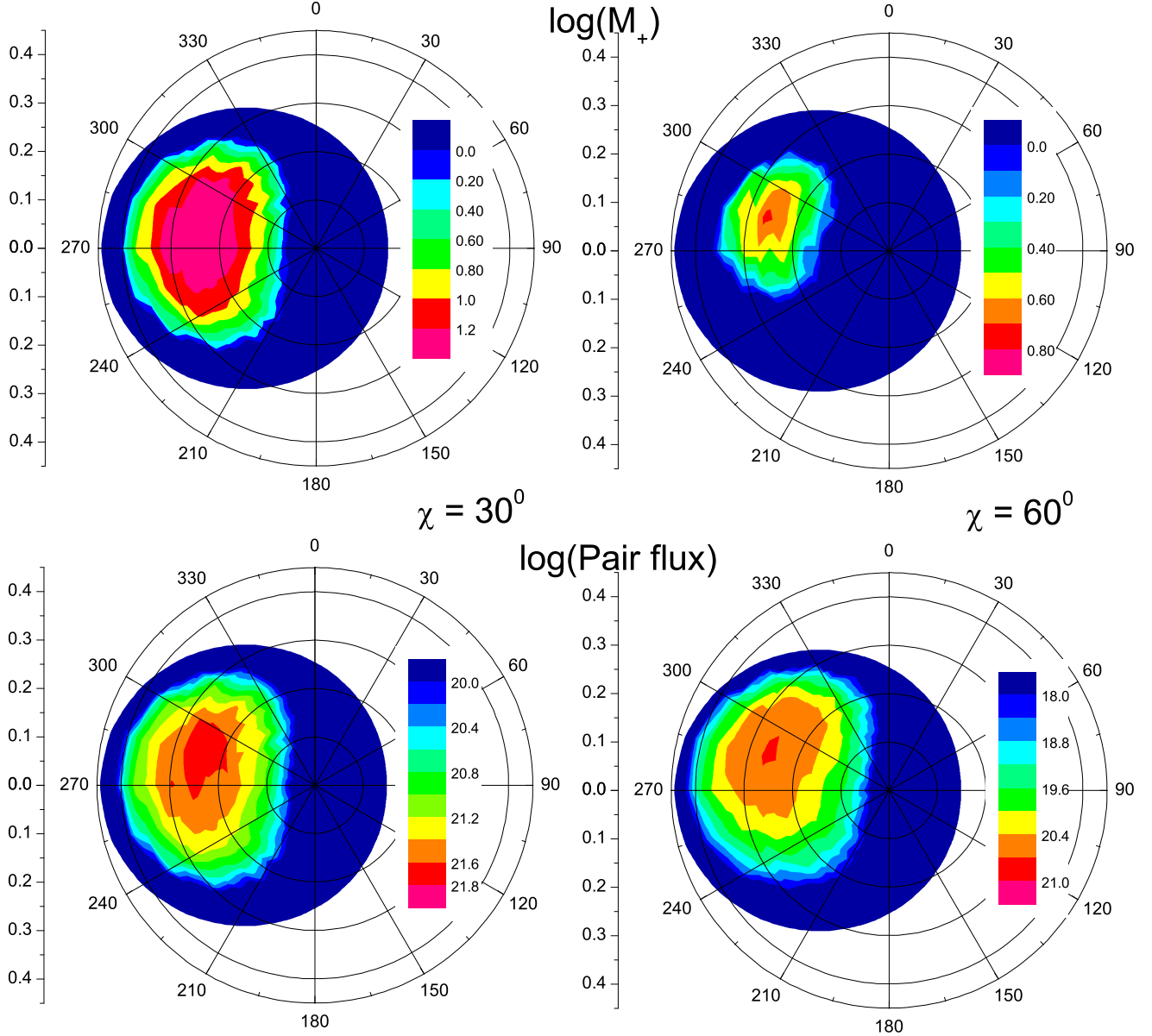}
\caption{Contours of log of pair multiplicity (top) and pair flux (bottom in units of $\rm s^{-1}\,cm^{-2}$) 
as a function of radial distance from the magnetic axis (in units of neutron star radius) and magnetic azimuth for $P = 3$ ms, $B_0 = 4 \times 10^8$ G and $\chi = 45^\circ$,
for $\varepsilon = 0.4$ and different values of inclination angle $\chi$. }    
\end{figure}

\newpage 
\begin{figure}
\hspace{-2.0cm}
\includegraphics[width=210mm]{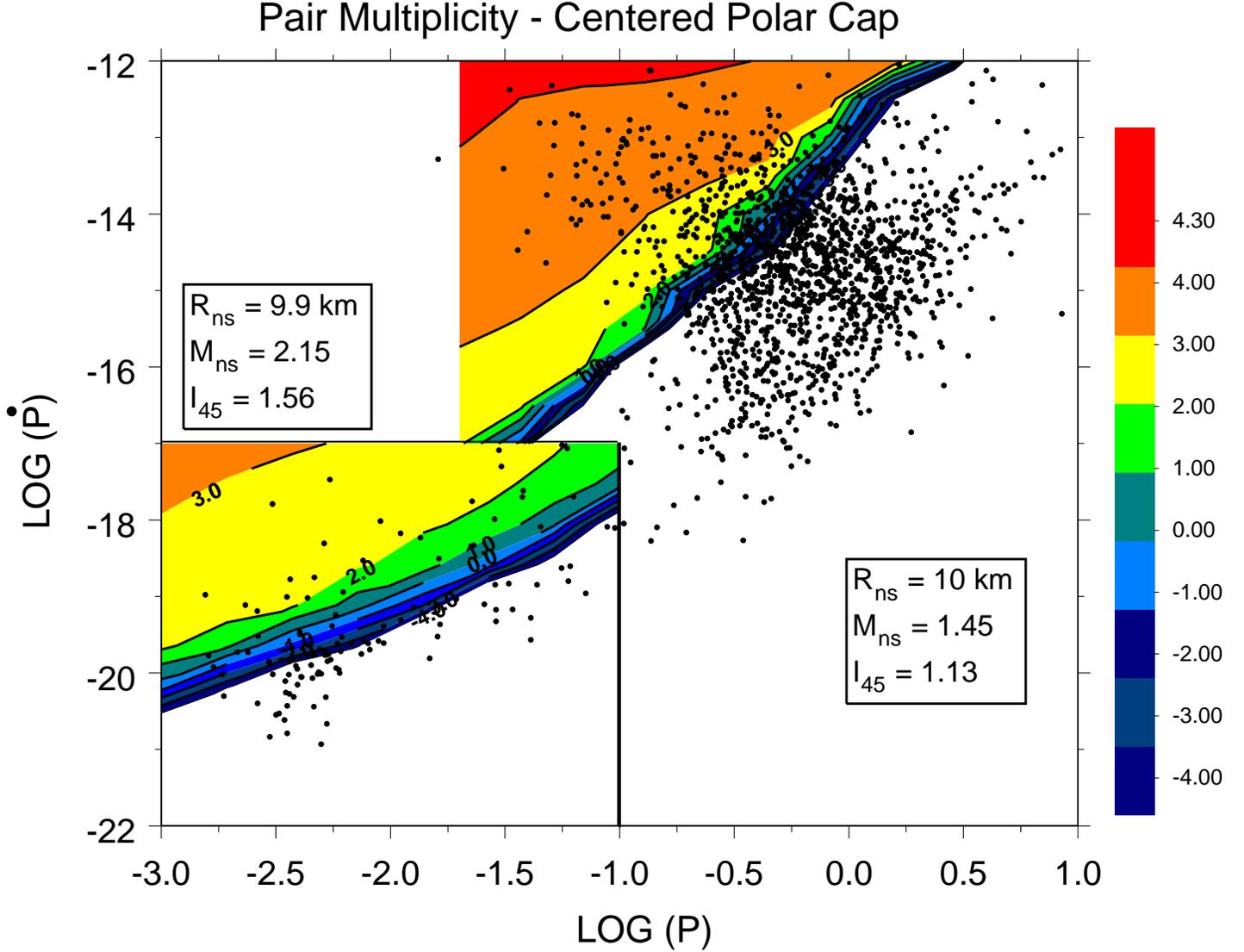}
\caption{Contours of log of peak pair multiplicity as a function pulsar period $P$ and period derivative ($\dot P$), for a centered PC and inclination angle $\chi = 60^\circ$ for normal 
pulsars and $\chi = 45^\circ$ for MSPs.  The NS radius $R_{\rm ns}$, mass $M_{\rm ns}$ (in Solar mass units) and moment of inertia $I_{45} = I/10^{45}\,\rm g\, cm^2$ refer to different NS  equations of state used for normal and millisecond pulsars and are described in the text.  Radio pulsars with measured $\dot P$ from the ATNF catalog (Manchester et al. 2005, http://www.atnf.csiro.au/research/pulsar/psrcat) are plotted as black dots. }    
\end{figure}

\newpage 
\begin{figure}
\hspace{-2.0cm}
\includegraphics[width=210mm]{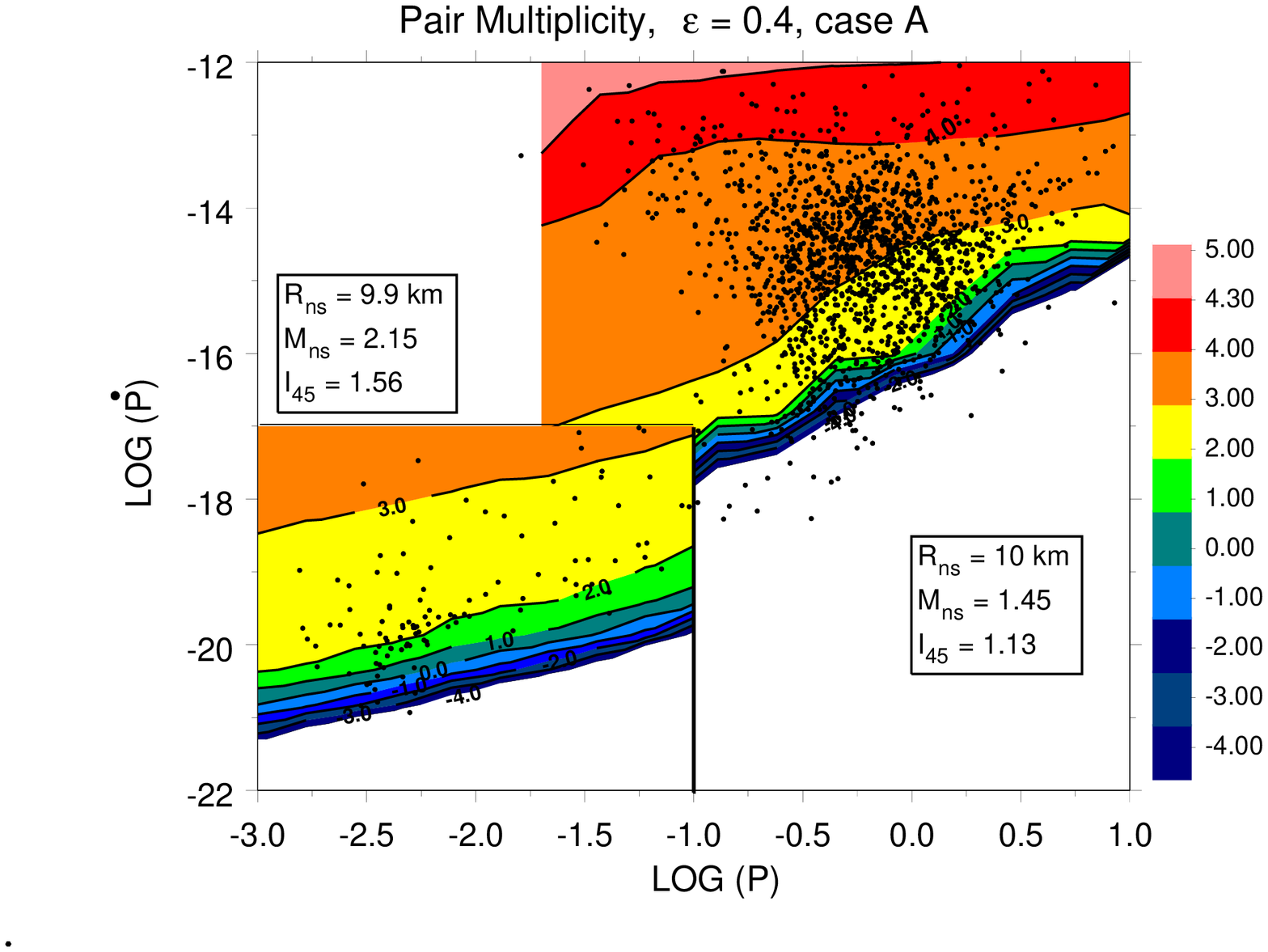}
\caption{Same as Figure 6 but for a symmetrically offset PC with $\varepsilon = 0.4$}    
\end{figure}

\begin{figure}
\hspace{-2.0cm}
\includegraphics[width=210mm]{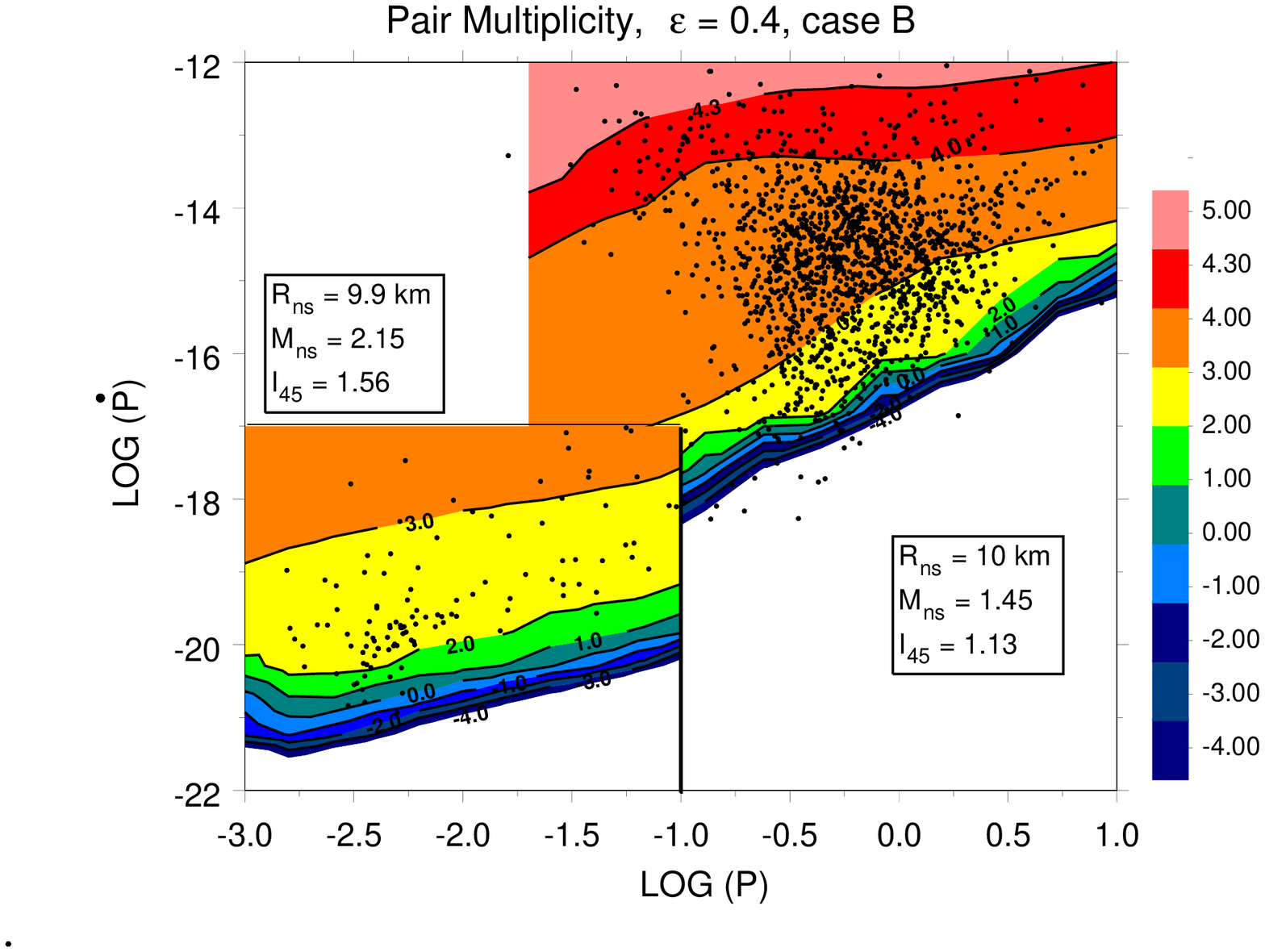}
\caption{Same as Figure 6 but for a asymmetrically offset PC with $\varepsilon = 0.4$}    
\end{figure}

\begin{figure}
\hspace{-2.0cm}
\includegraphics[width=210mm]{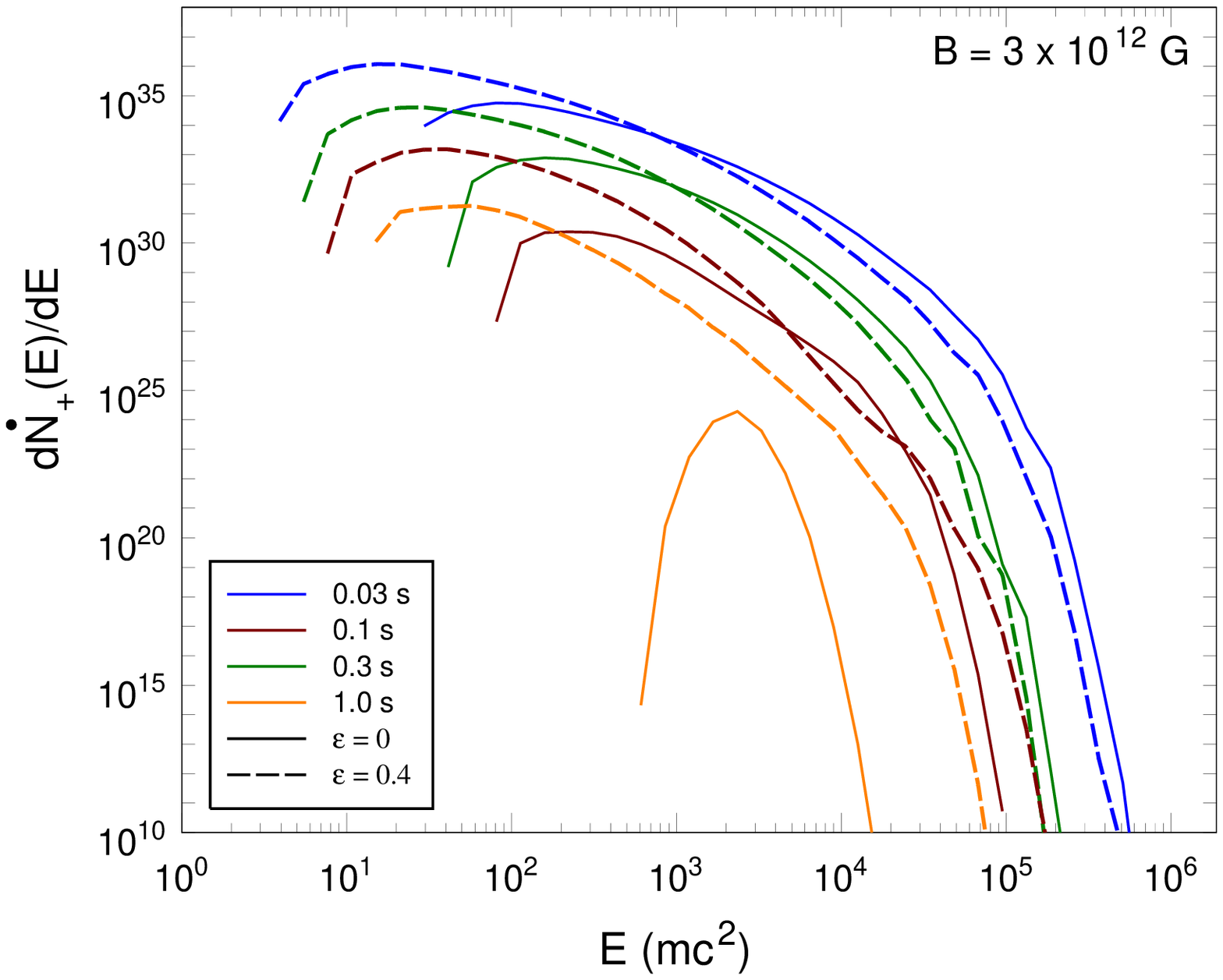}
\caption{ Spectra of pairs (pairs/(s $\rm mc^2$)) from one polar cap, for a surface magnetic field strength $B_0 = 3 \times 10^{12}$ G, different periods and two different 
degrees of offset $\varepsilon$.}   
\end{figure}

\newpage 
\begin{figure}
\hspace{-2.0cm}
\includegraphics[width=210mm]{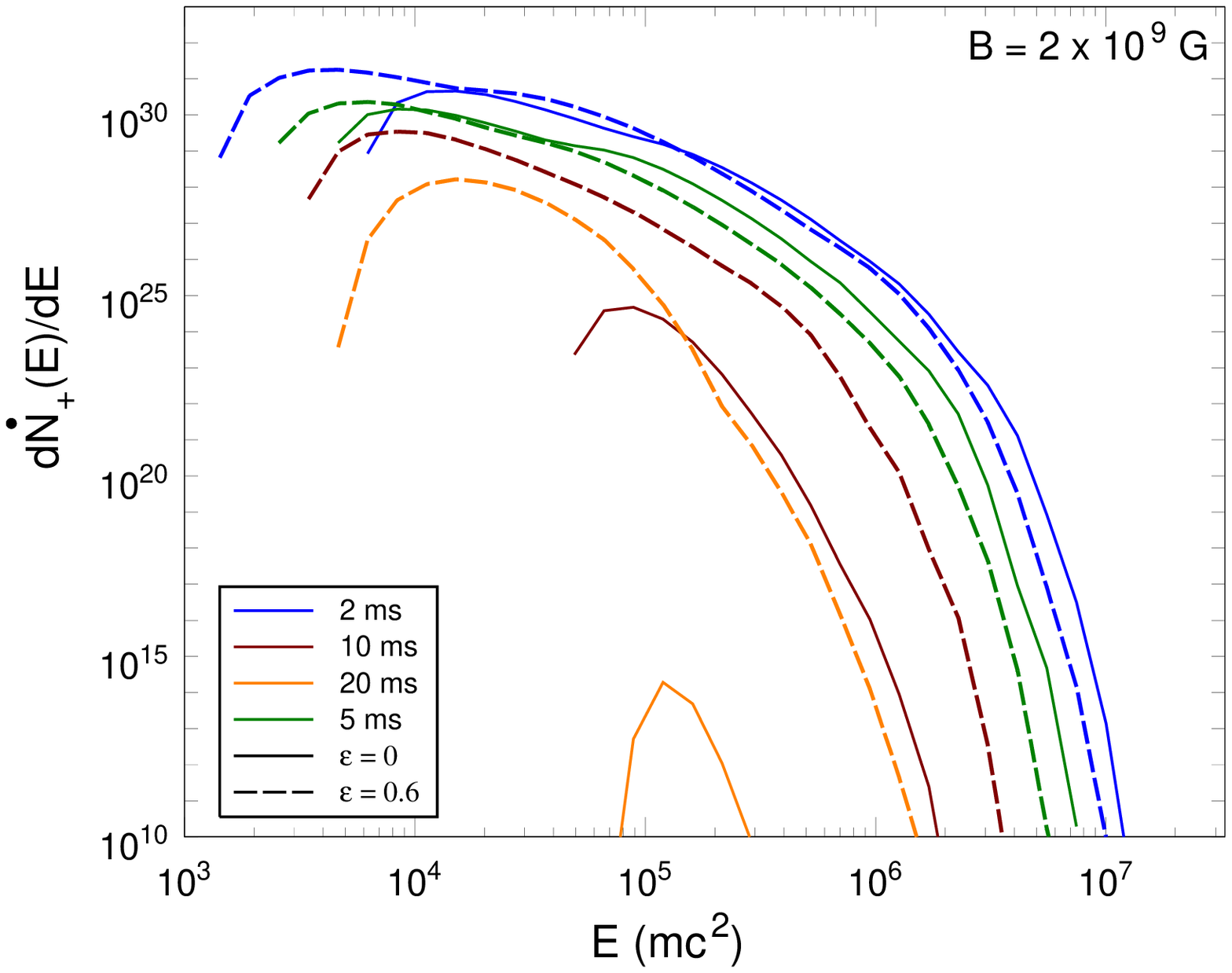}
\caption{
Spectra of pairs (pairs/(s $mc^2$)) from one polar cap, for a surface magnetic field strength $B_0 = 2 \times 10^{9}$ G, different periods and two different 
degrees of offset $\varepsilon$.}    
\end{figure}

\newpage 
\begin{figure}
\hspace{-2.0cm}
\includegraphics[width=210mm]{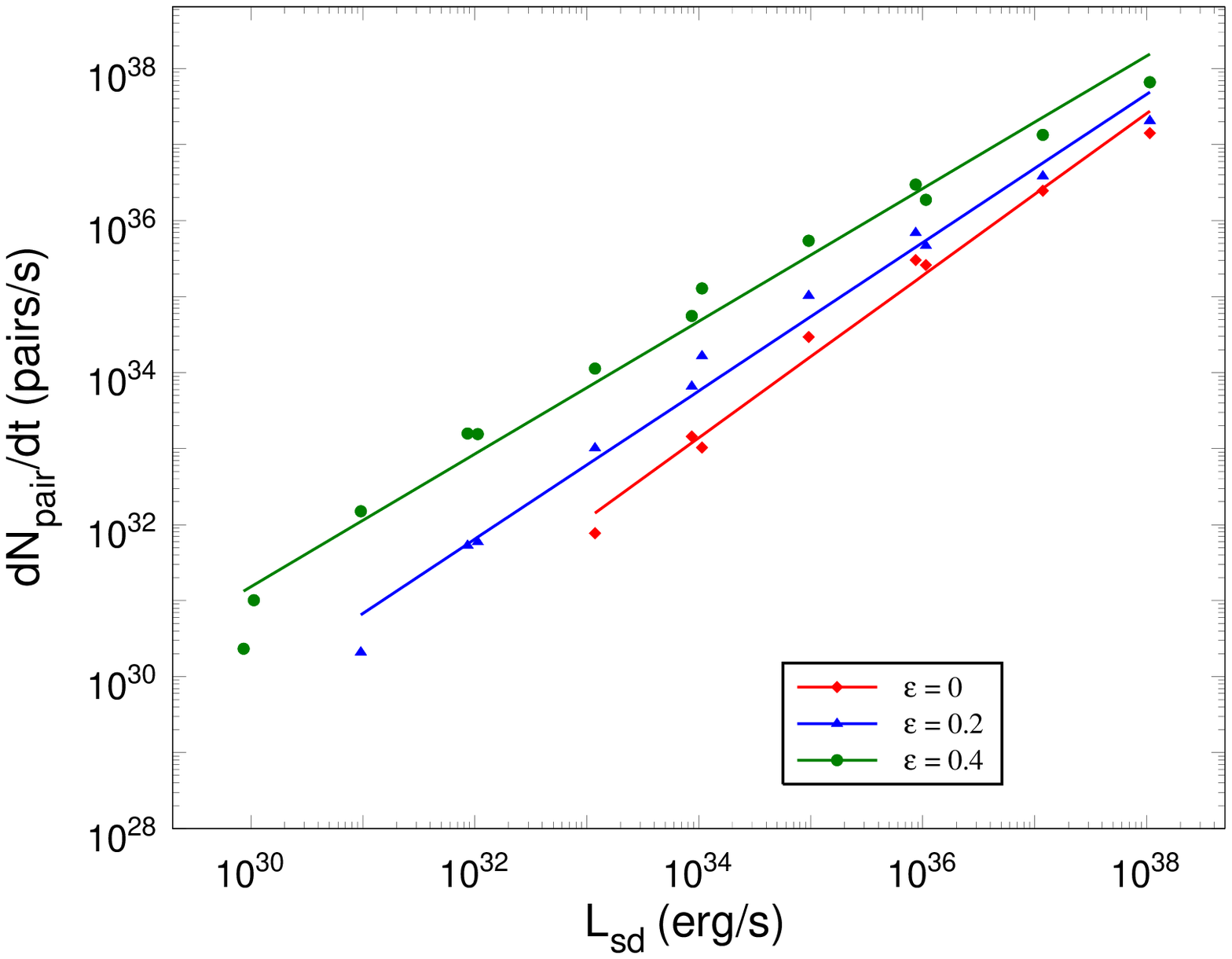}
\caption{Total pair flux (in units of pairs $\rm s^{-1}$) from each polar cap as a function of spin down luminosity $L_{\rm sd}$ for non-recycled pulsars, 
for different values of offset parameter $\varepsilon$.}    
\end{figure}

\newpage 
\begin{figure}
\hspace{-2.0cm}
\includegraphics[width=210mm]{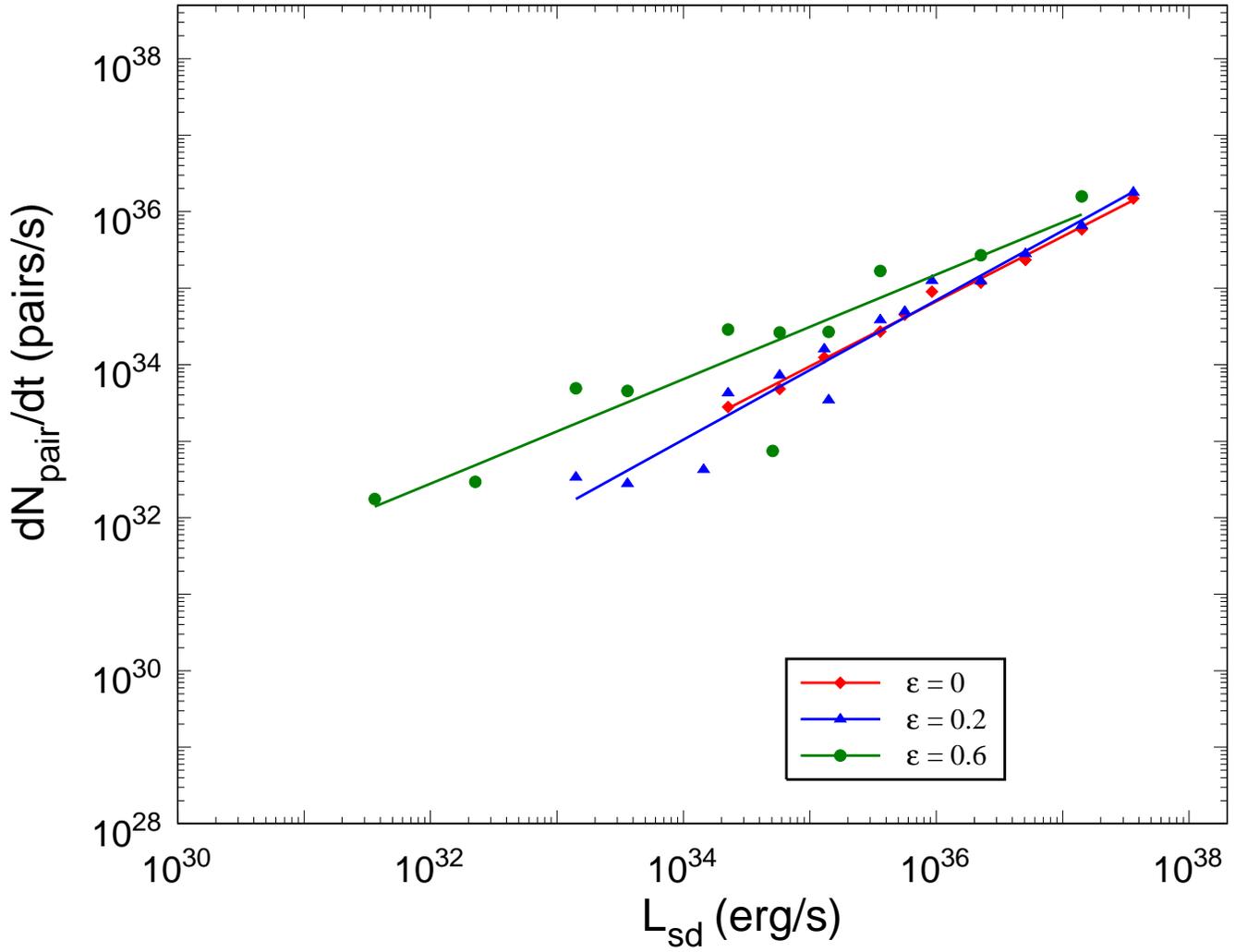}
\caption{Total pair flux (in units of pairs $\rm s^{-1}$) from each polar cap as a function of spin down luminosity $L_{\rm sd}$ for millisecond pulsars, 
for different values of offset parameter $\varepsilon$.}    
\end{figure}

\newpage 
\begin{figure}
\hspace{-2.0cm}
\includegraphics[width=210mm]{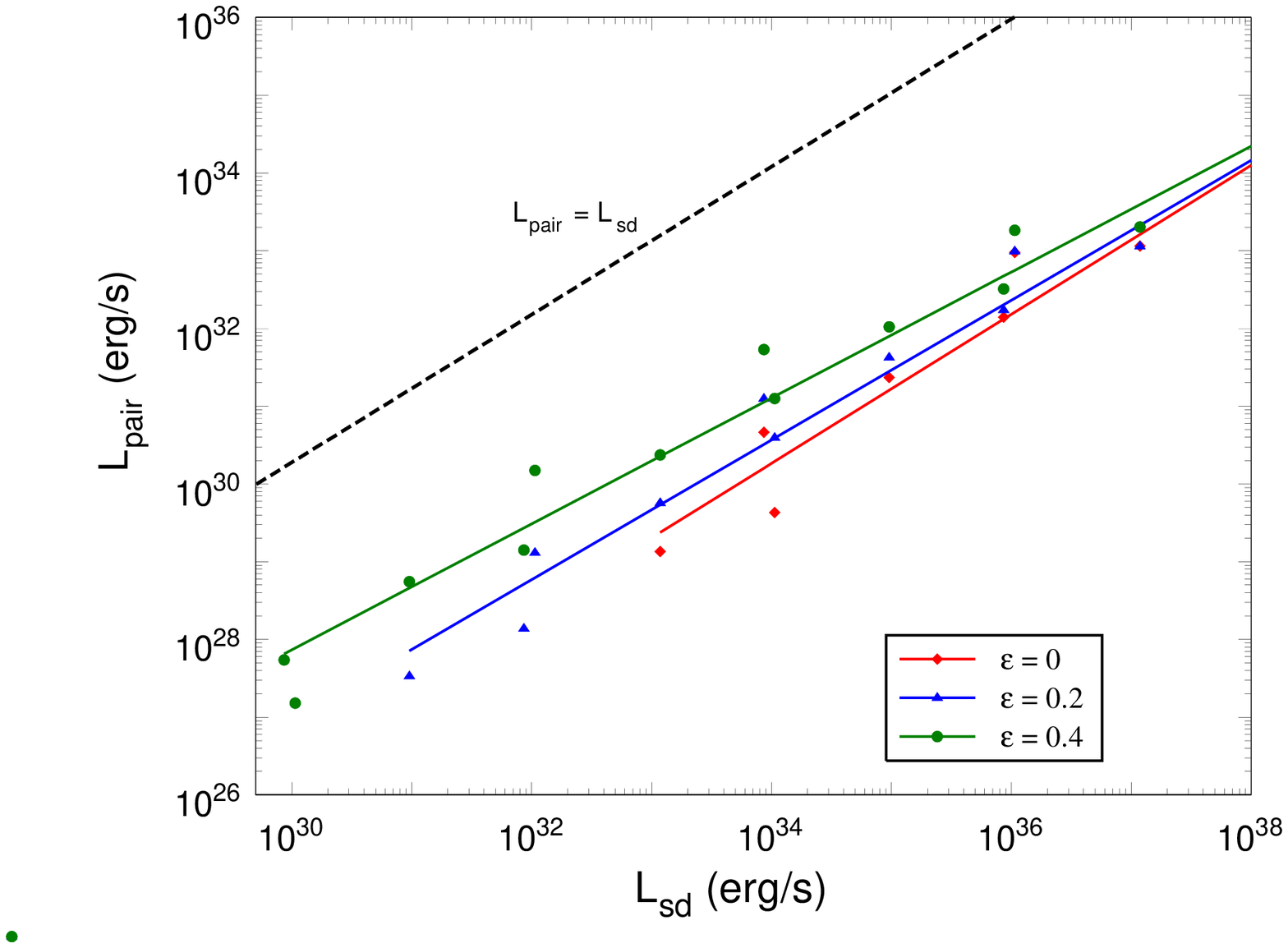}
\caption{Total pair luminosity (in units of erg $\rm s^{-1}$) from each polar cap as a function of spin down luminosity $L_{\rm sd}$ for non-recycled pulsars, 
for different values of offset parameter $\varepsilon$.}    
\end{figure}

\newpage 
\begin{figure}
\hspace{-2.0cm}
\includegraphics[width=210mm]{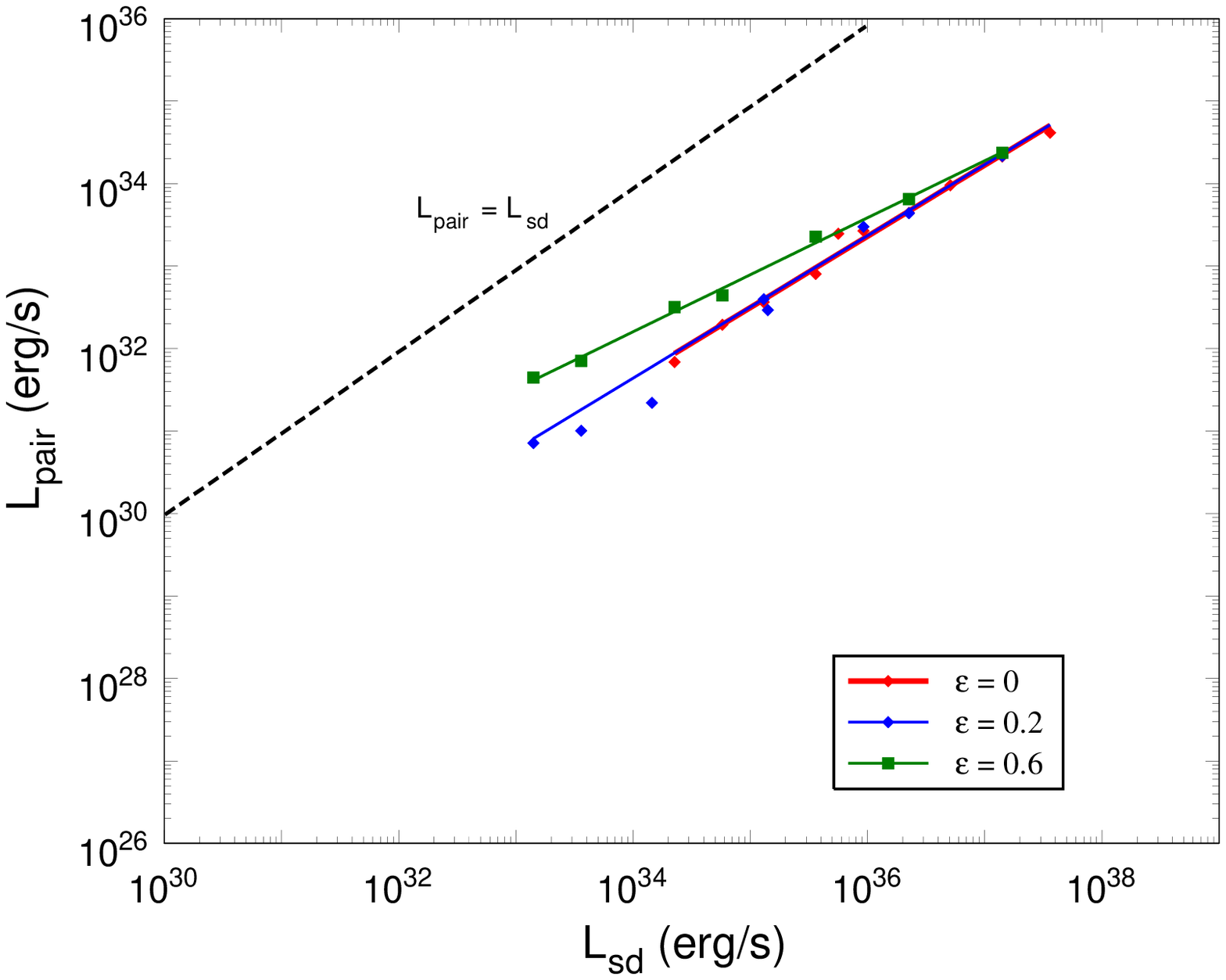}
\caption{Total pair luminosity (in units of erg $\rm s^{-1}$) from each polar cap as a function of spin down luminosity $L_{\rm sd}$ for millisecond pulsars, 
for different values of offset parameter $\varepsilon$.}    
\end{figure}

\newpage 
\begin{figure}
\hspace{-2.0cm}
\includegraphics[width=210mm]{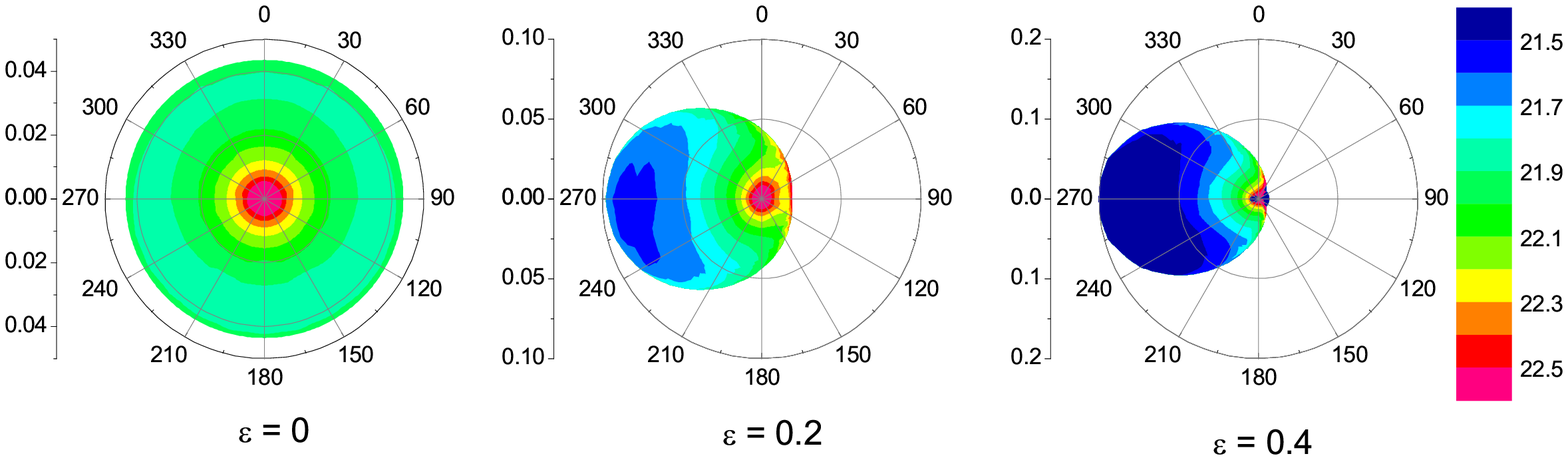}
\caption{Contours of log of positron heating flux (in units of $\rm erg\,s^{-1}\,cm^{-2}$) as a function of radial distance from the magnetic axis 
(in units of neutron star radius) and magnetic azimuth for $P = 0.1$ s, $B_0 = 3 \times 10^{12}$ G and $\chi = 60^\circ$,  for different values of offset parameter $\varepsilon$.  }    
\end{figure}

\begin{figure}
\hspace{-2.0cm}
\includegraphics[width=210mm]{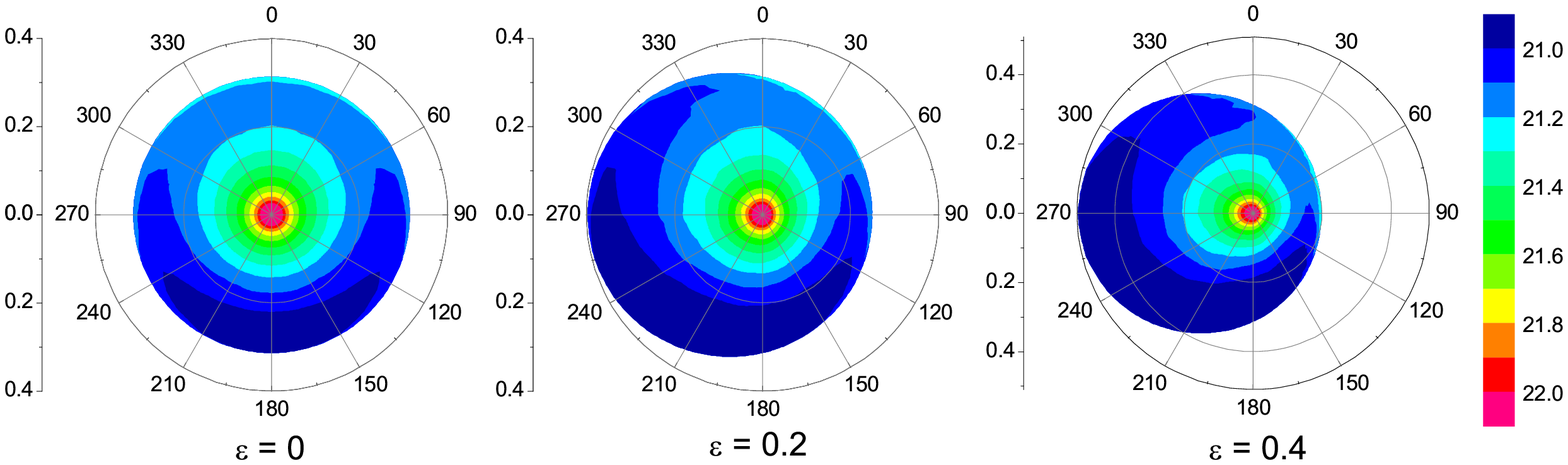}
\caption{Contours of log of positron heating flux (in units of $\rm erg\,s^{-1}\,cm^{-2}$) as a function of radial distance from the magnetic axis 
(in units of neutron star radius) and magnetic azimuth for $P = 2$ ms, $B_0 = 2 \times 10^{9}$ G and $\chi = 45^\circ$,  for different values of offset parameter $\varepsilon$.  }    
\end{figure}

\newpage 
\begin{figure}
\hspace{-2.0cm}
\includegraphics[width=210mm]{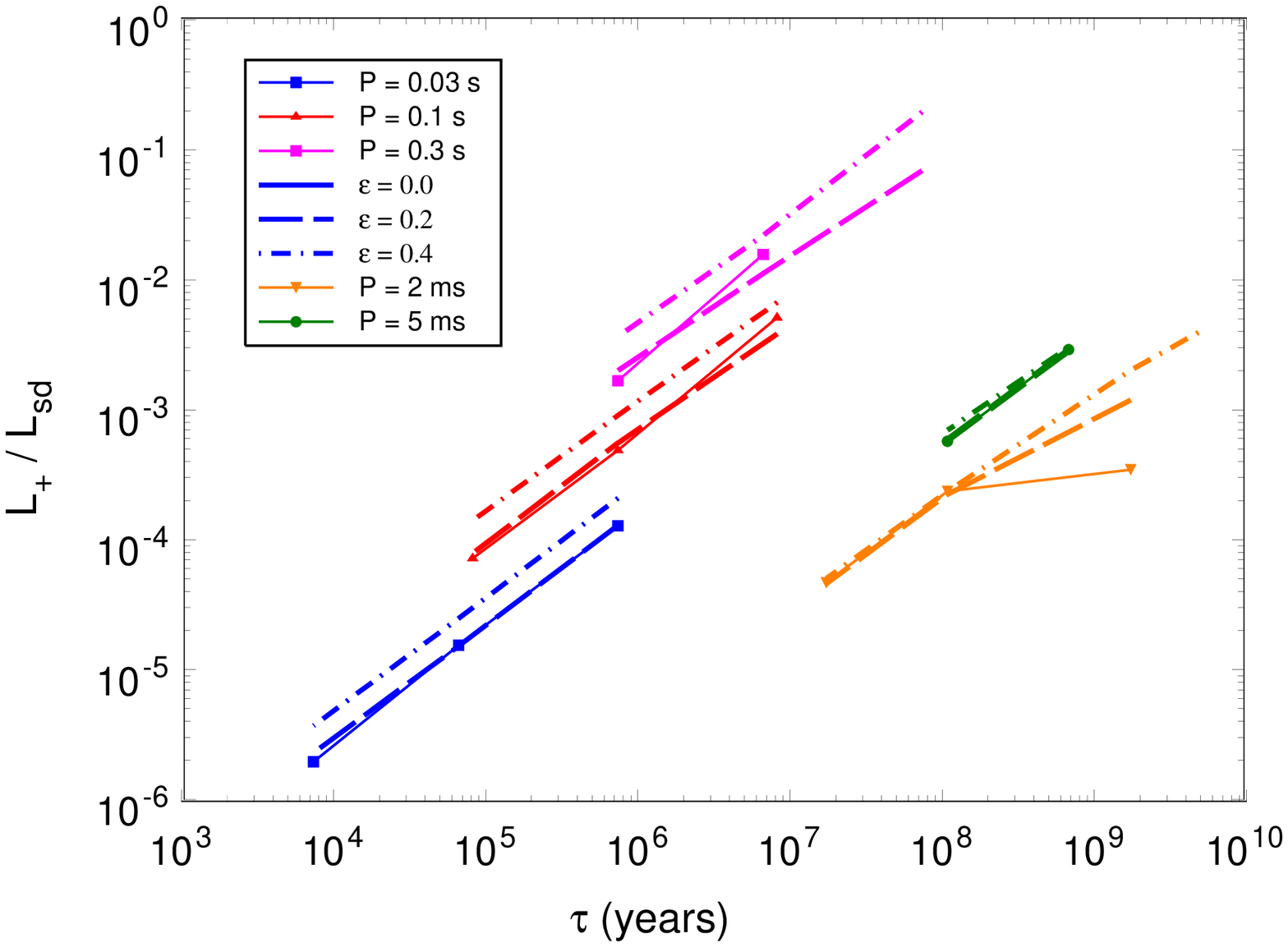}
\caption{Positron heating luminosity as a fraction of spin down luminosity $L_{\rm sd}$ vs. pulsar characteristic age $\tau$, for different pulsar periods and values of offset 
parameter $\varepsilon$}    
\end{figure}



\clearpage
\begin{references}
\reference{}
Abdo, A. A. et al. 2009, Science, 325, 848.
\reference{} 
Abdo, A. A. et al. 2010, ApJ, 712, 957
\reference{}
Abdo, A. A. et al. 2010b, ApJS, 187, 460. 
\reference{}
Arendt \& Eilek, J. A. 1998, (arXiv:astro-ph/9801257) 
\reference{}
Arons, J. 1981, ApJ, 248, 1099.
\reference{}
Arons, J. 1983, ApJ, 266, 215.
\reference{}
Arons, J. 1996, Astron. Astrophys., 120, 49.
\reference{}
Arons, J. 1997, in Neutron Stars and Pulsars: Thirty Years after the Discovery, eds. N. Shibazaki, N. Kawai, S. Shibata and T. Kifune, University Academy Press, Inc., Tokyo, Japan, p. 339.
\reference{}
Arons, J. \& Scharlemann, E. T. 1979, ApJ, 231, 854
\reference{}
Asseo, E. \& Khechinashvili,  2002, MNRAS, 334, 743.
\reference{}
Bai, X-N. \& Spitkovsky, A. 2010, ApJ, 715, 1282.
\reference{}
Baring, M. G. \& Harding, A. K. 2001, ApJ, 547, 929.
\reference{}
Bogdanov, S. , Rybicki, G. B. \& Grindlay, J. E. 2007, ApJ, 670, 668.
\reference{}
Bogdanov, S. \& Grindlay, J. E. 2009, ApJ, 703, 2259.
\reference{}
Cheng, K.~S., Ho, C., \& Ruderman, M.~A. 1986, ApJ, 300, 500.
\reference{}
Contopoulos, I., Kazanas, D. \& Fendt, C. 1999, ApJ, 511, 351.
\reference{}
Daugherty, J. K. \& Harding, A. K. 1982, ApJ, 257, 603.
\reference{}
Daugherty, J. K. \& Harding, A. K. 1983, ApJ, 273, 761.
\reference{}
Daugherty, J. K. \& Harding, A. K. 1996, ApJ, 458, 278.
\reference{}
De Jager, O. C., A. K. Harding, Michelson, P. F. , Nolan, P. L., Sreekumar, P. \& Thompson, D. J. 1996, ApJ, 457, 253.
\reference{}
DeCesar, M. E., Harding, A. K., Miller, M. C., Kalapotharakos, C. \& Contopoulos, I. 2011, 2011 Fermi Symposium Proceedings, 
eConf C110509 (arXiv:astro-ph/0349176).
\reference{}
De Luca, A., Caraveo, P. A, Mereghetti, S., Negroni, M. \& Bignami, G. 2005, ApJ, 623, 1051. 
\reference{}
Demorest, P. B., Pennucci, T.,  Ransom, S. M.,  Roberts, M. S. E. \& Hessels, J. W. T. 2010, Nature, 467, 1081.
\reference{}
Dermer, C. D. 1990, ApJ, 360, 197.
\reference{}
Deutsch, A. J., 1955, Ann. d'Astrophys., 18, 1.
\reference{}
Dyks, J. \& Harding, A. K. 2004, ApJ, 614, 869.
\reference{}
Dyks, J. \& Rudak, B. 2003, ApJ, 598, 1201.
\reference{}
Friedman, J. L., Ipser, J. R., \& Parker, L. 1986, ApJ, 304, 115.
\reference{}
Harding, A. K. 1981, ApJ, 245, 267.
\reference{}
Harding, A.~K., Baring, M.~G. \& Gonthier, P.~L. 1997, ApJ, 476, 246.
\reference{}
Harding, A.~K., \& Muslimov, A.~G. 1998, ApJ, 508, 328 [HM98].
\reference{}
Harding, A.~K., \& Muslimov, A.~G. 2001, ApJ, 556, 987 [HM01].
\reference{}
Harding, A.~K., \& Muslimov, A.~G. 2002, ApJ, 568, 862.
\reference{}
Harding, A.~K., Muslimov, A.~G. \& Zhang, B. 2002, ApJ, 576, 366.
\reference{}
Harding, A. K., J. V. Stern, J. Dyks \& M. Frackowiak, 2008, ApJ, 680, 1376.
\reference{}
Harding, A.~K., \& Muslimov, A.~G. 2011, ApJ, 726, L10 [HM11].
\reference{}
Harding, A. K., DeCesar, M. E., Miller, M. C., Kalapotharakos, C. \& Contopoulos, I. 2011, 2011 Fermi Symposium Proceedings, 
eConf C110509 (arXiv:astro-ph/0349902).
\reference{}
Harding, A. K. \& Preece, R. 1987, ApJ, 319, 939.
\reference{}
Hibschman, J. A. \& Arons, J. 2001, ApJ, 560, 871.
\reference{}
Hirotani, K. 2008, Open Astronomy (arXiv:0809.1283)
\reference{}
Kalapotharakos, K., Kazanas, D., Harding, A. K. \& Contopoulos, I. 2011, ApJ, submitted. 
\reference{}
Kantor, E. M. \& Tsygan, A. I. 2003, Astronomy Reports, 47, 613.
\reference{}
Kantor, E. M. \& Tsygan, A. I. 2004, Astronomy Reports, 48, 1029.
\reference{}
Contopoulos, I., Kazanas, D. \& Fendt, C.  1999, ApJ, 511, 351.
\reference{}
Lamb, F. K.  et al, 2009, ApJ, 706, 417.
\reference{}
Lattimer, J. M.  \& Prakash, M. 2007, Phys. Rev., 442, 109.
\reference{}
Luo, Q.; Shibata, S.; Melrose, D. B. 2000, MNRAS, 318, 943.
\reference{}
Manchester, R. N., Hobbs, G. B., Teoh, A. \& Hobbs, M. 2005, Astron. J., 129, 1993
\reference{}
Medin, Z. \& Lai, D. 2010, MNRAS, 406, 1379.
\reference{}
Mitra, D. \& Rankin, J. 2011, ApJ, 727, 92.
\reference{}
Muslimov, A. G. \& Tsygan, A. I. 1992, MNRAS, 255, 61 
\reference{}
Muslimov, A. G. \& Harding, A. K.  2004, ApJ, 606, 1143
\reference{}
Romani, R.~W. 1996, ApJj, 470, 469.
\reference{}
Ruderman, M. 1991, ApJ, 366, p. 261. 
\reference{}
Ruderman, M.A. \& Sutherland, P. G. 1975, ApJ, 196, 51 
\reference{}
Sokolov, A. A. \& Ternov, I. M. 1968, Radiation from Relativistic Electrons, ed. C. W. Kilmister (AIP: New York)
\reference{}
Spitkovsky, A. ApJ, 648, L51 (2006)
\reference{}
Li, J., Spitkovsky, A. \& Tchekhovskoy, A. 2011, ApJ, submitted (arXiv:1107.0979).
\reference{}
Takata, J.; Chang, H.-K.; Cheng, K. S. 2007, ApJ, 656, 1044.
 \reference{}
Thompson, D. J., Harding, A.K., Hersen, W. \& Ulmer, M.P. 1997, in Proc.
   of the {\it 4th Compton Symposium}, ed. C.D. Dermer, M.S. Strickman \& J.D. Kurfess         
   (AIP 410: New York), 39.
\reference{}
Timokhin, A. MNRAS, 36, 1055 (2006)
\reference{}
\reference{}
Timokhin, A. \& Arons, J. 2011, ApJ, in prep.
\reference{}
Tsai, W. Y \& Erber, T. 1974, Phys. Rev. D, 10, 492.
\reference{}
Venter, C.; Harding, A. K.; Guillemot, L. 2009, ApJ, 707, 800.
\reference{}
Vink, J, Mamba, A. \& Tamazaki, R. 2011, ApJ, 727, 131.
\reference{}
Zavlin, V. E. 2007, Astrophysics Space Sci., 308, 297.
\reference{}
Zhang, B., Harding, A. K. \& Muslimov, A. G.  2000, ApJ, 531, L135.


\end{references}
\end{document}